\documentclass[lettersize,journal]{IEEEtran}

\usepackage[table,xcdraw]{xcolor}

\usepackage{multirow}
\usepackage{tabulary}
\usepackage{pdflscape}
\usepackage{longtable}
\usepackage{makecell}
\usepackage{lscape}
\usepackage{amsmath} 
\usepackage{tikz}
\usepackage{amssymb}
\usepackage{tablefootnote}
\usepackage{array}
\usepackage{float}
\usepackage{url}
\usepackage{graphicx}
\usepackage[colorlinks=true, allcolors=blue,hyperfootnotes=true]{hyperref}

\newcommand{\AK}[1]{{\color{cyan} Arina: #1}}
\newcommand{\OG}[1]{{\color{blue} Olga: #1}}

\begin{document}

\title{Secure Software Development Methodologies:\\ A Multivocal Literature Review}

\author{Arina~Kudriavtseva, Olga~Gadyatskaya
\IEEEcompsocitemizethanks{\IEEEcompsocthanksitem A. Kudriavtseva and O. Gadyatskaya are with Leiden Institute of Advanced Computer Science, Leiden University, The Netherlands \protect\\
}\thanks{In submission}}

\maketitle

\begin{abstract}

In recent years, the number of cyber attacks has grown rapidly. An effective way to reduce the attack surface and protect software is adoption of methodologies that apply security at each step of the software development lifecycle. While different methodologies have been proposed to address software security, recent research shows an increase in the number of vulnerabilities in software and data breaches. Therefore, the security practices incorporated in secure software development methodologies require investigation.

This paper provides an overview of security practices involved in 28 secure software development methodologies from industry, government, and academia. To achieve this goal, we distributed the security practices among the software development lifecycle stages. We also investigated auxiliary (non-technical) practices, such as organizational, behavioral, legal, policy, and governance aspects that are incorporated in the secure software development methodologies. Furthermore, we explored methods used to provide evidence of the effectiveness of the methodologies. Finally, we present the gaps that require attention in the scientific community. 

The results of our survey may assist researchers and organizations to better understand the existing security practices integrated into the secure software development methodologies. In addition, our bridge between ``technical'' and ``non-technical'' worlds may be useful for non-technical specialists who investigate software security. Moreover, exploring the gaps that we found in current research may help improve security in software development and produce software with fewer number of vulnerabilities.
\end{abstract}


\section{Introduction}\label{sec:introduction}

\IEEEPARstart{A}{ccording} to Common Vulnerabilities and Exposures database of MITRE~\cite{cvemitre}, the number of reported vulnerabilities has been increasing since 1999. Forbes reported that ``every minute, \$2,900,000 is lost to cyber crime and top companies pay \$25 per minute due to cyber security breaches''~\cite{forbes}. The escalating risk of cyber attacks has led to the concept of ``shifting security left'', which emphasizes performing security practices early in the software development process, rather than leaving them for testing or post-deployment phases. This ``shifting left'' concept has prompted organisations to implement a secure software development lifecycle. 

In the early 2000s, personal computers connected to the Internet became more widespread~\cite{microsoftaboutsdl}. This trend provided attackers with opportunities to target remote machines, leading to a surge in self-propagating malware. Existing security practices in the industry at that time were inadequate~\cite{howard2006security}, necessitating a fundamentally different approach to protect organisations from malicious software.

The first publications systematically studying how to build secure software emerged in 2001~\cite{mcgraw2004software,wing2003call,mcgraw2002building,mcgraw2003ground,leblanc2002writing,viega2001building}. From 2004 onward, organizations began integrating security processes into the software development life cycle (SDLC). For example, in 2004, Microsoft finalized the Security Development Lifecycle (SDL) and incorporated it into their software development processes. Since then, many companies and organisations have developed their own approaches to produce secure software~\cite{sap2020,citrix2021,cisco2021}. The increasing number of these approaches calls for a systematic investigation to identify their similarities and differences.

This literature review aims to investigate and summarize the security practices involved in each step of established secure software development methodologies. As the target audience of these methodologies is organisations engaged in software development, a multivocal study covering methodologies from industry, government organizations and academic research is most appropriate. In our survey, we map the security practices used in the methodologies according to the SDLC stages, as is customary for such methodologies~\cite{howard2006security}. It is intriguing that there is a plethora of methodologies focused on the same end goal, with new ones regularly emerging. 

While several surveys~\cite{de2009secure, davis2005secure, gregoire2007secure, fonseca2013survey, williamssecure, nunez2020preventive, ramirez2020survey} have investigated and compared existing secure software development methodologies (SSDMs), to the best of our knowledge, ours is the most comprehensive, covering 28 SSDMs from industry, government and academia that were issued between 2004 and 2022. We begin by comparing the methodologies to each other based on included security practices, such as threat modeling or static security analysis. Reflecting the emerging understanding in the field that software security is not solely a technical pursuit, but also a socio-technical one~\cite{pirzadeh2010human,mokhberi2021sok}, we pay special attention to auxiliary (non-technical) practices that support software security. At the same time, we examine supporting evidence that the studied SSDMs effectively enhance software security by reviewing validation studies (including validation reports with the methodologies themselves, if available). Finally, we identify gaps in the literature and propose new research directions that address these gaps.

To summarize, the contributions of our research are:
\begin{itemize}
    \item we systematized security practices involved in 28 SSDMs;
    \item our systematization covers practices integrated in the SDLC and auxiliary (non-technical) practices that support software security;
    \item we systematize the existing evaluation approaches for secure software development methodologies;
    \item we report on the discovered gaps that require more attention in the research community.
\end{itemize}

\section{Research methodology}\label{sec:methodology}

In this paper, we conducted a multivocal literature review following the guidelines proposed by Garousi, Felderer and M{\"a}ntyl{\"a}~\cite{garousi2019guidelines} that consolidates both academic and industry sources.

\subsection{Study focus}
Numerous synonyms are used to describe approaches that incorporate security practices at each step of the SDLC. For example, the publications have used the following terms: \textit{secure software development process}~\cite{de2009secure,gregoire2007secure}, \textit{secure software development lifecycles}~\cite{sap2020,cisco2021,citrix2021,isaca,apvrille2005secure}, \textit{secure development lifecycle}~\cite{microsoftsdl,ge}, \textit{secure software development framework}~\cite{nist218}, \textit{security-by-design framework}~\cite{csa2017}, \textit{framework}~\cite{bsa2020,alkussayer2010isdf,khan2011secure,chatterjee2013framework}, \textit{secure software development}~\cite{griponssd2015}, \textit{guidelines}~\cite{malaysia2015}, \textit{model}~\cite{sodiya2006towards,daud2010secure}, \textit{methodology}~\cite{farhan2018methodology}.

In our survey, we utilize the term \textit{secure software development methodologies} (SSDMs), to denote a collection of high-level secure software development practices integrated into the SDLC. We consider the term \textit{methodology} to be synonymous with \emph{guideline, lifecycle, model, and framework}.

\noindent \textbf{Scope:}
This survey specifically focuses on secure software development methodologies that incorporate security practices in every phase of the SDLC. Therefore, software assurance maturity models~\cite{owaspsamm,bsimm}
and methods that concentrate on a specific stage of the SDLC~\cite{weider2012towards,van2017design,meland2008secure,4359475} lie beyond the scope of this research. Additionally, we aim to survey general SSDMs while excluding those that are applicable only to a specific technology (e.g., mobile, IoT, cloud).

\subsection{Related concepts}

During our literature search on SSDMs, we encountered the terms \textit{DevSecOps} and \textit{application security}, which are occasionally used to describe security practices in software development processes. Since these terms are relevant to our research topic, we incorporated them into our search criteria. However, we found that these terms did not yield many relevant sources.

\subsubsection{DevSecOps}\label{DevSecOps}

DevOps, an abbreviation for development and operation, refers to the integration of development and operation teams. 
In 2012, MacDonald and Head~\cite{macdonald2016devsecops} from Gartner discussed the necessity of incorporating security into DevOps, thus introducing the term DevSecOps. Presently, one of the challenges faced by organizations adopting DevOps is ensuring secure software delivery~\cite{rajapakse2022challenges}. 

There have been numerous academic studies (e.g.,~\cite{rajapakse2022challenges,mohan2016secdevops,farroha2014framework, schneider2015security, rahman2016software, cash2016managed, de2014continuous, myrbakken2017devsecops, ahmed2019integrating, macdonald2016devsecops}) and industry experience reports (e.g.,~\cite{synopsis2020,AWSdevsecops}) on DevSecOps. However, these studies do not provide a comprehensive description of the security practices involved in each phase of the SDLC as required by inclusion criterion IN-1 (Table~\ref{tab:criteria}). Therefore, we have excluded DevSecOps studies from our research.

\subsubsection{Application Security}\label{AppSec}

McGraw~\cite{mcgraw2006software} posited that software security is about building security in, while application security is about protecting the software in a reactive way after development is complete. Similarly, Payne~\cite{payne2010} studied the challenges of application security initiatives that are involved after software has been developed. The author proposed a proactive approach to implementing application security in software development projects. However, as Payne's approach~\cite{payne2010} does not cover all stages of the SDLC, as required by the inclusion criterion IN-1 (Table~\ref{tab:criteria}), we do not include it in our study. 

Chakraborty~\cite{chakraborty2016} from Synopsis offered a viewpoint on comparing application security with software security. According to Chakraborty, application security is a subset of software security and focuses on post-deployment issues (such as patching, IP filtering, and post-deployment security tests), while software security addresses pre-deployment issues. 

ISO/IEC 27034~\cite{ISO27034} provides guidance on security techniques for application security. The guidance focuses on specifying, designing, and implementing security controls throughout the entire SDLC. 
However, as we could not find a free version of the standard, we did not include it in our research. 

To summarize, we found no application security methodology in academic papers, but rather literature focused on specific technologies, such as mobile, web, and cloud application security. Since our interest lies in general methodologies that incorporate security practice in each phase of the SDLC, we did not include this literature in our survey. We found only one industry publication~\cite{rebit2020} that presents an application security framework meeting our criteria, which we discuss in Section~\ref{sec:industry}.

\subsubsection{Software assurance maturity models}\label{maturity}

Software assurance maturity models enable organizations to assess the capabilities and maturity of their software security practices within the SDLC. These models are developed based on software security surveys, allowing organizations to compare their practices with those of peers who have already implemented software security initiatives~\cite{owaspappsec}. The security practices outlined in maturity models are structured into multiple maturity levels. Lower maturity levels encompass relatively easier-to-implement security practices within their corresponding categories. It is not mandatory for organizations to achieve the highest maturity level in each category. Instead, the level of maturity should be determined based on the specific needs of the organization. 

We identified two software maturity models: OWASP Software Assurance Maturity Model (SAMM)~\cite{owaspsamm} and Building Security In Maturity Model (BSIMM)~\cite{bsimm}. These models consist of software security frameworks designed to organize security activities and assess security initiatives. While our research focuses on general security practices rather than evaluating progress (maturity), the security practices included in these software security frameworks align with our inclusion criteria. Therefore, we discuss these frameworks in Section~\ref{sec:industry}.

\noindent In this study, we aim to address the following research questions:

\begin{itemize}
\item[\textbf{RQ1}] What are the existing general approaches for secure software development?
\item[\textbf{RQ2}] What are the \emph{auxiliary} steps that these methodologies use besides the security practices integrated into the usual SDLC?
\item[\textbf{RQ3}] How have these approaches been evaluated in terms of their effectiveness?
\end{itemize}

\subsection{Literature search methodology}

\textbf{Search strategy}: To conduct this literature review, we employed Google Scholar~\footnote{\url{https://scholar.google.nl/}} to identify relevant academic papers, while Google Search~\footnote{\url{https://www.google.nl/}} was utilized to locate relevant grey literature, such as blogs, white papers, reports, government documents.

\noindent \textbf{Search terms}: To search for secure software development methodologies, we utilized the following search strings:

\begin{enumerate}
    \item \texttt{"Secure" AND (("Software" AND ("Engineering" OR ("Development" AND ("Methodology" OR "Framework" OR "Model" OR "Standard" OR "Lifecycle")))) OR (("Systems" OR "Software") AND ("by Design" OR "design")))}
    \item \texttt{("DevSecOps")AND ("Methodology" OR "Framework" OR "Model" OR "Standard" OR "Lifecycle")}
    \item \texttt{("Application Security" OR "AppSec")AND ("Methodology" OR "Framework" OR "Model" OR "Standard" OR "Lifecycle")}
\end{enumerate}

\noindent \textbf{Study selection}: Upon receiving the initial search results, we excluded irrelevant literature using inclusion and exclusion criteria (Table~\ref{tab:criteria}).

\begin{table}[t]
\caption{Inclusion and exclusion criteria for literature}
\label{tab:criteria}
\begin{tabulary}{0.5\textwidth}{p{1cm}p{7cm}}
\hline
 & Inclusion criteria \\ \hline
IN-1 & Literature discussing the methodologies which incorporate security practices into each phase of the SDLC \\
IN-2 & Literature written in English \\
IN-3 & Full text is available (either it is free or it is included in our academic subscription)\\
IN-4 & Include only the five first pages on Google Search\\ \hline
 & Exclusion criteria\\ \hline
EX-1 & Literature discussing security only in a particular phase of the SDLC\\
EX-2 & Evident advertisement of a vendor or a product\\
EX-3 & Literature discussing a methodology for a specific technology (IoT, web applications, etc.)\\ 
EX-4 & Agile methodology does not map security practices to agile requirements (every-sprint, bucket, one-time) \\ \hline
\end{tabulary}
\end{table}

\begin{figure}[t!]
\begin{center}
    \includegraphics[width=3.5in]{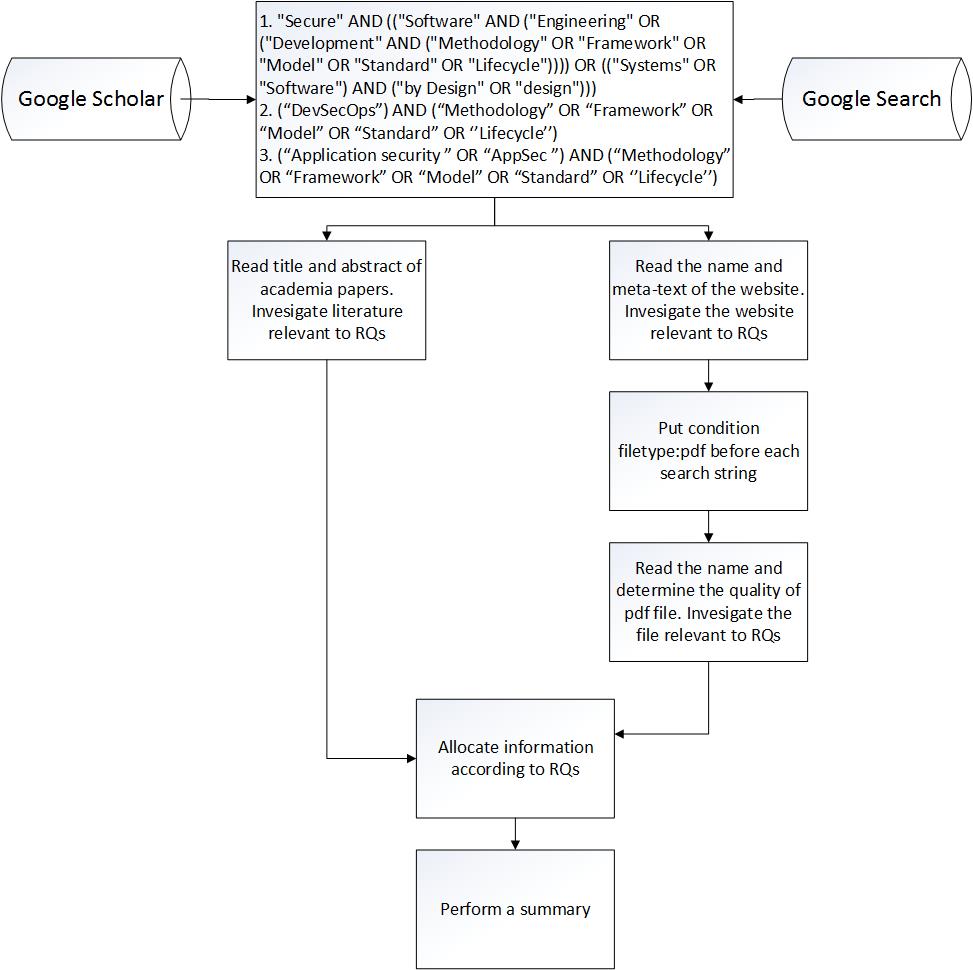}
    \caption{Literature search process}
    \label{pic:search}
\end{center}
\end{figure}

\noindent \textbf{Search procedure}: We identified the initial set of the methodologies using the search process outlined in Fig.~\ref{pic:search}. For academic and grey literature, we employed two distinct search procedures.

To select relevant academic papers, initially, we reviewed the titles and abstracts of the articles. We then applied the inclusion and exclusion criteria, selecting only the articles relevant to the topic of the study. Subsequently, we thoroughly read the full text of the selected literature and organized the information based on the RQs. 

For the grey literature search, our process comprised two stages. In the first stage, we conducted web search using predefined search strings. In the second stage, we expanded our search to include grey literature in the form of \texttt{pdf} documents. To this end, we added the search condition \texttt{filetype:pdf} before each search string. To determine the relevance of the grey literature, we evaluated the title and meta-text provided by Google Search. Similar to academic literature, we applied the inclusion and exclusion criteria to filter the grey literature. Afterwards, we carefully examined the full text of the relevant grey literature and utilized the AACODS checklist~\cite{grey} to assess the quality of the grey literature. Subsequently, we extracted data in accordance with the RQs. 

To identify additional academic and grey literature, we employed backward and forward snowballing techniques as recommended by Wohlin~\cite{wohlin2014guidelines}. Backward snowballing involved identifying new papers by examining the citations of the papers we were analyzing. Forward snowballing, on the other hand, involved identifying new papers through the reference lists of the analyzed papers.

Lastly, we cross-validated our findings with the list of existing international secure software engineering initiatives compiled by ENISA~\cite{enisa2011} in 2011. 

\section{Secure Software Development Methodologies}\label{sec:SSDM}
Based on the search procedure and the inclusion and exclusion criteria defined in the previous section, we have identified 28 SSDMs published between 2004 and 2022. In this section, we provide a brief summary of these methodologies.

\begin{table*}[t!]
\centering
\caption{Secure software development methodologies ordered chronologically}
\label{tab:allmethodologies}
  \begin{tabulary}{1\textwidth}{CLC}
\hline
The source of methodology & Name & Year of publication \\ \hline
\multirow{15}{*}{Industry} & Microsoft Software Development Life Cycle (SDL)~\cite{howard2006security,microsoftsdl} & 2006 \\
 & McGraw’s Secure Software Development Lifecycle Process~\cite{mcgraw2006software, verdon2004risk} & 2006 \\
 & Comprehensive, Lightweight Application Security Process (CLASP)~\cite{web:lang:stats} & 2006 \\
 & Microsoft SDL version 5.2 for Agile Development~\cite{whitepapermicrosoft} & 2012 \\
 & Software Assurance Forum for Excellence in Code (SAFECode)~\cite{safecode2018} & 2018 \\
 & Building Secure and Reliable Systems~\cite{adkins2020building} & 2020 \\
 & BSA framework~\cite{bsa2020} & 2020 \\
 & The Secure Software Development Lifecycle at SAP~\cite{sap2020} & 2020 \\
 & ReBIT Application Security Framework~\cite{rebit2020} & 2020\\
 & OWASP Software Assurance Maturity Model~\cite{owaspsamm} & 2020\\
 & Cisco Secure Development Lifecycle~\cite{cisco2021} & 2021 \\
 & Citrix Security Development Lifecycle~\cite{citrix2021} & 2021 \\
 & Building Security in Maturity Model~\cite{bsimm12} & 2021\\
 & GE Secure Development Lifecycle~\cite{ge} & 2022 \\ \hline
 \multirow{3}{*}{Government}  &  Grip on Secure Software Development~\cite{griponssd2015} & 2015 \\
 & CSA Singapore Security-by-Design~\cite{csa2017} & 2017 \\
 & SSDLC guidelines Malaysia~\cite{malaysia2015} & 2020 \\
 & Security in SDLC Romania~\cite{isaca} & 2021 \\
 & NIST 800-218~\cite{nist218} & 2022 \\ 
 & NIST 800-160~\cite{NIST2016v1r1}  & 2022 \\ \hline
\multirow{9}{*}{Academia} & Secure Coding: Building Security into the Software Development Life Cycle~\cite{jones2004secure} & 2004 \\
 & Secure Software Development Life Cycle Process~\cite{apvrille2005secure} & 2005 \\
 & The Secure Software Development Model (SSDM)~\cite{sodiya2006towards} & 2006 \\
 & The Integrated Security Development Framework (ISDF)~\cite{alkussayer2010isdf} & 2010 \\
 & Secure Software Development Model: A Guide for Secure Software Life Cycle~\cite{daud2010secure} & 2010 \\
 & Secure Software Development: a Prescriptive Framework~\cite{khan2011secure} & 2011 \\
 & Framework for Development of Secure Software~\cite{chatterjee2013framework} & 2013 \\
 & Methodology for Enhancing Software Security During Development Processes~\cite{farhan2018methodology} & 2018 \\
 \hline
 \end{tabulary}
\end{table*}

\begin{figure*}[h!]
\centering
    \includegraphics[width=0.8\textwidth]{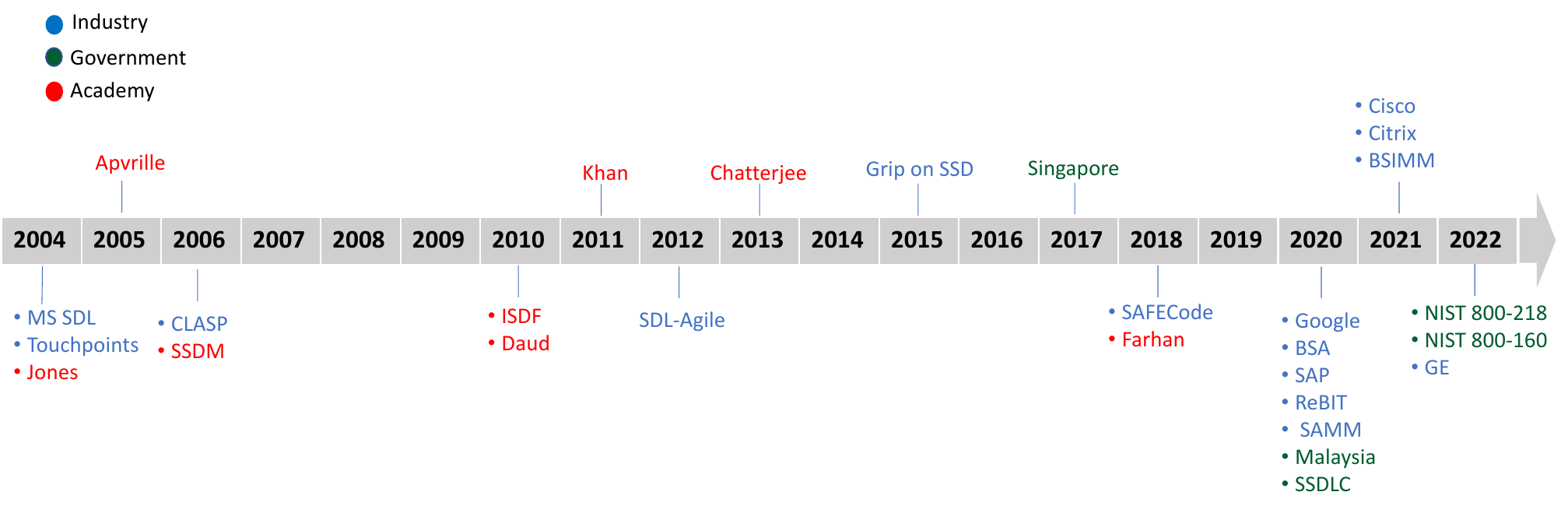}
    \caption{The timeline}
    \label{pic:timeline}
\end{figure*}

The timeline depicting the publication dates of the methodologies is shown in Fig.~\ref{pic:timeline}. It is important to note that the publication date may not necessarily coincide with the date of the methodology's introduction. For example, SDL~\cite{howard2006security} was incorporated into Microsoft's software development process in 2004, but the methodology was officially published in 2006, two years later.

\subsection{The structure of the SDLC}
In order to discuss the security practices incorporated in the analyzed methodologies, it is important to first define the structure of the SDLC. The SDLC encompasses crucial phases involved in software development. Although there are various approaches to SDLCs, we categorize all lifecycles into two main groups: (1) Waterfall, which follows a sequential development model, and (2) Agile, which adopts an iterative approach. When authors of a methodology do not specify the SDLC category to which their security practices belong (which is the case for the majority of methodologies), we consider such methodologies to fall under the Waterfall category. Only when authors explicitly refer to Agile, do we classify a methodology accordingly.

Different presentations of the SDLC stages exist. In our research, we combined all the SDLC stages from the methodologies to create a unified set of stages. The only exception is the \textit{test plans} phase in McGraw Touchpoints~\cite{mcgraw2006software}. Since this phase is unique to Touchpoints and not consistent with other methodologies, we combined it with the design phase. The final set of SDLC stages and their brief explanation are as follows:
\begin{itemize}
    \item project inception, or planning phase provides a high-level overview of the project goals and requirements, along with preliminary activities for software development; 
    \item analysis and requirement phase involves creating and maintaining software requirements;
    \item architectural and detailed design phase ``identifies the major components of the system and the communication between these components''~\cite{sommerville2006};
    \item implementation phase involves writing a program based on the requirements;
    \item verification and testing phase ensures that the software meets the requirements and fulfills the customer's expectations;
    \item release and maintenance phase includes activities related to release preparation, deployment, and post-production maintenance;
    \item disposal phase involves activities to retire the software.
\end{itemize}

During the investigation of security practices, we discovered that certain practices cannot be associated with specific stages of the SDLC. Some practices are project-wide, covering all stages of the SDLC, while others are organization-wide, applicable to all projects within a company. As a result, we divided all security practices into three categories: (1) organization-wide, (2) practices that cover all stages of the SDLC, and (3) practices specific to a particular project. To determine the placement of each practice, we referred to the framework mentioned in the respective texts. In cases where the applicability of a practice to a stage was not explicitly stated, the authors discussed and made a joint decision on where to position it.

Some authors emphasize that their security approaches are process-agnostic, and their security practices do not explicitly map to specific stages of the SDLC~\cite{mcgraw2006software, griponssd2015}. While conducting this literature review, we provided our interpretation of the methodologies, aiming to adhere as closely as possible to the authors' ideas. However, if the author did not explain a specific security practice in the text, we chose not to include these practices in the table. For instance, in the ISDF methodology ~\cite{alkussayer2010isdf}, there is no explanation of the \textit{logging and tracing} practice in the coding phase.

Table~\ref{tab:waterfall} displays 26 SSDMs for Waterfall software development, sorted by publication date and classified by origin. The two methodologies applicable to Agile development are presented in Table~\ref{tab:agile}. In this table, bucket practices are marked in green, every-sprint practices are marked in red, and one-time practices are marked in blue. Both tables provide the (potentially abbreviated) names of the security practices for each studied methodology. If a practice's name corresponds to the column name, it is denoted with a checkmark $\checkmark$. To ensure clarity, we utilize the term SAST to refer to static analysis and source code analysis.

\onecolumn
\begin{landscape}
\tiny 
\renewcommand{\arraystretch}{0.8}
\setlength{\tabcolsep}{2pt}
\setlength\LTcapwidth{\textwidth} 


\end{landscape}

\twocolumn

Next, we briefly summarize the distinct aspects of each methodology.

\subsection{Industrial methodologies}\label{sec:industry}
\subsubsection{Microsoft Secure Development Lifecycle {SDL}}

Microsoft first introduced the integration of security and privacy considerations into all phases of SDLC in 2004. In 2006, Howard and Lipner released the methodology ``The security development lifecycle (SDL)''~\cite{howard2006security}. Microsoft emphasizes two secure practices: executive support and education and awareness, as these steps have been successful in reducing the number of code bugs~\cite{howard2006security}. The company ensures mandatory security training, exercises, and labs for all engineering staff. Microsoft also gives equal attention to secure design principles and secure coding best practices. At the time of the book's publication, the authors claimed that most of the SDL secure practices could also be incorporated into Agile software development. 

However, in 2012, Microsoft published a white paper~\cite{whitepapermicrosoft}, introducing a new version 5.2 of SDL with the addition of SDL for Agile development (SDL-Agile). The SDL-Agile methodology will be discussed in Section~\ref{sdlagile}. The main difference between Microsoft SDL 2006 and Microsoft SDL v.5.2 is the inclusion of privacy concerns. As a non-technical practice, privacy is considered in RQ2.

Compared to the version 5.2, in the latest version of Microsoft SDL~\cite{microsoftsdl}, Microsoft has added the following practices. In the inception stage, the \textit{metrics definition and compliance reporting} process was created to establish a minimum level of security quality before starting the project. Additionally, it is important to define and approve the list of tools for software development. The \textit{define and use cryptography standards} process was added to ensure the use of only approved encryption libraries during project development. The modern version of Microsoft SDL~\cite{microsoftsdl} is no longer includes the \textit{security response execution} process.

\subsubsection{McGraw Touchpoints}

In 2004, McGraw published a paper~\cite{mcgraw2004software} introducing the concept of touchpoints. He later expanded on this concept in his book ``Software Security: Building Security in''~\cite{mcgraw2006software} published in 2006, which built upon previous research~\cite{viega2001building} and~\cite{hoglund2004exploiting}. We classify this methodology as an industry approach because McGraw was affiliated with Cigital. McGraw describes software security as an ongoing process based on three pillars: (1) applied risk management, (2) software security best practices (touchpoints), and (3) knowledge~\cite{mcgraw2006software}. While touchpoints concentrate on security practices, knowledge management and risk management are integral parts of any software development project. 

Knowledge management plays a crucial role in training secure development staff on the most important security issues to increase awareness. It helps to establish the understanding that security is everyone's responsibility within the organization, including builders, operations personnel, administrators, users, and executives. In addition to education and awareness tasks, the knowledge pillar encompasses guidelines, principles, rules, and historical knowledge that can be applied throughout the SDLC. 

The touchpoints are directly linked to the stages of the Waterfall software development process. However, they can be applied regardless of software development approach used and can be cycled through multiple times as the software evolves. McGraw highlights two touchpoints as particularly critical: code review and architectural risk analysis. These two touchpoints are combined because addressing software security problems correctly requires both code review and architectural risk analysis. 

McGraw treats the test plans phase as a separate practice, referred to as the \textit{risk-based security preparation security} practice. This phase is based on the abuse case scenarios developed during the analysis and requirement phases and includes a set of constructive and destructive activities. In Table~\ref{tab:waterfall}, we position test plans within the design phase.

A notable aspect of the McGraw framework is that it incorporates ongoing external analysis (review) throughout all stages of the SDLC. This review is conducted by individuals outside the company. McGraw also emphasizes that risk analysis should be a continuous process throughout the requirement, design and testing phases rather than a single step. The results of the risk analysis guide the formulation of requirements and the planning of specific tests. Penetration testing is also emphasized as a continuous process, covering the verification and testing, release and maintenance phases to ensure the security of the system in its deployment environment.

\subsubsection{Comprehensive, Lightweight Application Security Process (CLASP)}

In 2006, Dan Graham published a set of processes ``Introduction to the CLASP Process''~\cite{graham2006introduction} to assist software development team in incorporating security considerations at the early stages of SDLC. Although the methodology was revised in 2013, the link to the updated document no longer functions. We therefore focused in the original document published in 2006.

CLASP (stands for Comprehensive, Lightweight Application Security Process) was later adopted by the OWASP consortium~\footnote{\url{https://owasp.org/}} and is recognized as a lightweight methodology suitable for small organizations with less stringent security requirements~\cite{de2009secure}. CLASP follows a role-based approach, where security practices are tailored to specific project roles.

One ongoing practice within CLASP that spans the project's lifecycle is the \textit{monitoring of security metrics}. This practice helps to measure the project's progress or the performance of the project team. Similar process, such as \textit{defining and using criteria for software security checks} In NIST 800-160 and the \textit{defining metrics and compliance reporting} in Microsoft SDL, address the same objective.

The \textit{identify user roles and resource capabilities} practice in CLASP involves mapping roles and their associated capabilities. This practice also considers the role of potential attackers. It shares similarities with the \textit{understanding adversaries} practice found in the methodology by Google~\cite{adkins2020building}. 

In addition to providing a comprehensive description of the best security practices, CLASP offers worksheets with coding guidelines to support the implementation of these security practices.

\subsubsection{SAFECode}

In 2018, the Software Assurance Forum for Excellence in Code (SAFECode) published the ``SAFECode Fundamental Practices for Secure Software Development: Essential Elements of a Secure Development Lifecycle Program''~\cite{safecode2018} to assist the industry in adopting security software development practices.  

Rather than duplicating security principles, SAFECode refers to the sources of these practices. For describing the security design principles, SAFECode refers to Saltzer and Schroeder principles ~\cite{saltzer1975protection}. The practices for threat modeling are detailed in a SAFECode white paper~\cite{safecode2017}, and for managing third-party components, SAFECode has published another white paper~\cite{thirdparty2017}.

SAFECode places significant emphasizes on the importance of \textit{planning the implementation and deployment} of secure development practices, considering it an integral part of any healthy organization. One of the key practices in the planning phase is \textit{creating the product development model and lifecycle}. The goal of this practice is to integrate security and non-security specialists within a single framework, reducing friction when introducing security practices into the lifecycle. 

An exceptional practice highlighted by SAFECode is \textit{application security control definition}. It involves identifying threats, assessing risks, defining security requirements, validating the implementation of security requirements, and ensuring compliance with policies.

Another important practice is \textit{standardizing identity and access management}, which encompasses mechanisms for authentication and authorization. This practice aligns with the BSA methodology~\cite{bsa2020}, which also emphasizes the significance of identity and access management.

SAFECode argues that when an organization plans to introduce new practices, it should consider activities that contribute to building a security-focused culture. Such activities may include learning from the experience of other organizations, analyzing past mistakes, and highlighting successful activities.

\subsubsection{The methodology by Google}

In 2020, Adkins et al. released the book ``Building Secure \& Reliable Systems''~\cite{adkins2020building}, which was a collaboration between O’Reilly and Google. The authors emphasize that although the book focuses on security, general approaches can also be applied to achieve privacy goals. For brevity, we refer to this methodology as the methodology by Google. 

The authors argue that in order to establish security requirements, it is important to understand and assess the motivation of potential attacker. In addition, it is crucial to consider potential risks from insiders. Further discussion on understanding adversaries' processes can be found in Section~\ref{sec:nontechnical}. 


From the authors' perspective, the term supply chain refers to the processes of writing, building, testing, and deploying software. To enhance the security of the software supply chain against insider threats, code review and automation are deemed crucial tactics. Moreover, automated systems can perform various steps in the supply chain, reducing human involvement and minimizing mistakes. The authors also advocate for the inclusion of binary provenance and verifiable builds to protect against adversaries. 

The book also focuses on disaster preparedness, response during a disaster, and recovery after a disaster. To ensure that system's resilience and continuity during a disaster, effective \textit{disaster planning} is essential. During security crises, crisis management plays a crucial role in enabling the system to withstand attacks.   According to the authors, crisis management involves detailed plans and effective communication, including an operational security (OpSec) plan.The OpSec plan determines which information needs to remain confidential and how the response should proceed without exposing the organization to further risks. After a significant security incident, the recovery phase aims to mitigate the attack and restore the system to its normal state while incorporating necessary improvements.

In conclusion, the authors highlight that all the security practices described in the book can be effective if a company has a culture of security and reliability. This aspect is further explored in Section~\ref{sec:nontechnical}.

\subsubsection{The BSA framework}

In 2020, the BSA foundation published a white paper ``The BSA Framework for Secure Software: A new approach to securing the software lifecycle''~\cite{bsa2020}, which we refer to as the BSA framework. The structure of the BSA framework is designed to be applicable for organizations of all sizes, including those working with the Internet of Things, Artificial Intelligence, and various software development methods, including DevOps.

In Table \ref{tab:waterfall}, secure development processes are documented throughout software development activities, covering all stages of the SDLC. However, security guidance for development and testing activities may consist of general practices applicable to all projects within a company. Additionally, gathering and documenting security requirements is an integral part of the analysis and requirements phase.

The \textit{supply chain} category includes practices related to third-party risk management, ensuring the protection of supply chain data. It also encompasses practices related to the implementation phase, such as ensuring software integrity, software identification, and ensuring proper usage of software.

The authors emphasize the significance of organizational processes and product security capabilities as essential components of secure software. Security capabilities encompass various technical aspects that should be considered during the software design phase. These capabilities include support for identity management and authentication, patchability, cryptographic services, authorization and access controls, logging, and error and exception handling.

Within the BSA framework, the term ``SDL Governance'' refers to building a culture of security within the organization. This involves establishing policies, standards, and metrics to promote a strong security posture. 

\subsubsection{The SAP methodology}

In 2020, SAP corporation published a white paper ``The Secure Software Development Lifecycle at SAP''~\cite{sap2020}, referred to as the SAP methodology. SAP places significant emphasis on the preparation stages defined in the ISO/IEC 27034-1 standard~\cite{ISO27034}.

The SAP methodology incorporates three types of risk assessment modeling to identify and analyze risks: product-level assessment, scenario-based assessment, and fast-track threat modeling. Following the \textit{risk assessment} practice, the next step is \textit{security planning}. This involves determining security and privacy requirements and implementing security controls to mitigate identified risks. The security controls are categorized into two groups: (1) security functions to enforce software security, and (2) measures taken by the product team to prevent vulnerabilities. In the SAP methodology, the analysis and requirements phase merge with the design phase, as there are no clear boundaries between them. This merged phase is reflected in Table \ref{tab:waterfall}.

Within SAP, the software development phase combines the design and implementation stages. Product teams employ secure design principles, secure programming techniques, libraries, and tools to ensure the security of the software. 

\subsubsection{ReBIT Application Security Framework}

In 2020, Reserve Bank Information Technology (ReBIT) published an application security framework as a guide for Chief Information Security Officer (CISO) to implement application security within the organizations.

The application security lifecycle consists of four main stages: (1) request for proposal, (2) development lifecycle, (3) production rollout, and (4) post deployment processes. The request for proposal phase is used to define the security requirements for the organization. These requirements may encompass various aspects, including secure design, secure deployment, security assessment, disaster recovery, secure use of open source, security compliance with policies and processes, and security for support and maintenance. This phase can be considered as the project planning phase. 

One remarkable practice within the ReBIT framework is vulnerability assessment and penetration testing (VAPT). The framework prescribes a set of minimum tests that must be performed, such as grey box and white box testing, web application testing, testing of underlying infrastructure for thick client applications, mobile application testing, Windows application testing, and handhold device application testing. 

\subsubsection{OWASP SAMM}

In 2020, OWASP published Software Assurance Maturity Model (SAMM)~\cite{owaspsamm} version 2.0, This model supports various software development methodologies, including Waterfall, Iterative, Agile, and DevOps. It categorizes 15 security practices into five groups aligned with business functions: (1) governance, (2) design, (3) implementation, (4) verification, and (5) operation. 

During the implementation phase, the SAMM framework emphasizes the importance of the \textit{defect management} practice, which involves tracking and analyzing security defects within a project. By utilizing the acquired information, organizations can effectively reduce the occurrence of new defects. 

In the operation domain, \textit{environment management} plays a crucial role in ensuring a secure environment. This process includes activities such as patching, updating, and configuration hardening after the software is released. It is worth noting that the BSA framework~\cite{bsa2020} also incorporates a similar practice called \textit{development environment}; however, BSA's focus is on protecting the development environment from security threats while software is being developed. 

In addition, SAMM offers organizations a self-assessment toolbox to measure their software assurance maturity performance.

\subsubsection{The Cisco methodology}

In 2021, Cisco published a white paper ``Secure Development Lifecycle''~\cite{cisco2021}, which outlines their secure-by-design philosophy. This approach, referred to as the Cisco methodology, emphasizes the importance of the planning phase in incorporating defense-in-depth techniques. Given Cisco's primary focus on cloud-based technologies, they also prioritize the security planning of these technologies, adhering to industry certifications such as SOC 2 Type II and ISO 27001.

The developing phase at Cisco includes internal security training programs designed to enhance the engineers' knowledge of security practices. This ongoing process is categorized under the education and awareness phase in Table~\ref{tab:waterfall}.

During the launch phase, Cisco places significant emphasis on security readiness to ensure products are prepared for customer use, including thorough checks of critical security and privacy controls. Additionally, the company maintains a channel known as the Product Security Incident Response Team that facilitates communication and collaboration with customers in order to address critical security risks effectively.

\subsubsection{The Citrix methodology}

In 2021, Citrix published the ``Citrix Security Development Lifecycle''~\cite{citrix2021}. We will refer to this as to the Citrix methodology.

Citrix places a strong emphasis on both internal and external engagement. For example, they have established a Red Team responsible for simulating attacks on projects throughout the year. Additionally, Citrix engages external companies to conduct security assessment and penetration testing. The Citrix Product Security Engineering team also conducts regular penetration tests. In addition, the company actively participates in the Bug Bounty program, allowing researchers to identify vulnerabilities in their products. By combining the findings from the Red Team work, external assessment, Bug Bounty program, and Product Security Engineering team, Citrix establishes the foundation for their security remediation programs.

The Citrix framework addresses supply chain security by not only managing third-party components but also by analysing, tracking, and testing components within the CI/CD (continuous integration, continuous delivery) pipeline. 

\subsubsection{BSIMM}

The latest version of Building Security In Maturity Model (BSIMM) was published in 2021~\cite{bsimm12}. This model is the result of analyzing gathered data on the security practices adopted by different organizations to address software security problems. The underlying structure of BSIMM is a software security framework consisting of 12 security practices. These practices are categorized into four domains: (1) governance, (2) intelligence, (3) SSDL Touchpoints, (4) deployment. Altogether, these practices encompass 112 security activities that are classified into three levels of maturity. As our research focuses on the underlying framework of BSIMM, we will not delve into the maturity models.

During our investigation, we identified a similarity between the \textit{knowledge} pillar of McGraw Touchpoints~\cite{mcgraw2006software} and the \textit{intelligence} domain of BSIMM. Both aim to gather and share knowledge within the organization, which can be applied at various stages of the SDLC for specific projects. The intelligence domain of BSIMM comprises three security practices: (1) attack models, (2) security features \& design, (3) standards \& requirements. These practices are considered organizational-wide as they contribute to the accumulation of corporate knowledge used for implementing software security activities throughout the organization~\cite{bsimm12}.

The touchpoints of BSIMM are associated with specific SDLC processes and include (1) architecture analysis, (2) code review, and (3) security testing.

The \textit{deployment} domain encompasses (1) penetration testing, (2) software environment, and (3) configuration management and vulnerability management. \textit{Software environment} includes configuration documentation, code signing, and change management.

According to BSIMM, the ten most commonly observed security activities in organizations are as follows: implementing lifecycle instrumentation and using it to define governance, ensuring basic host and network security measures are in place, identifying PII obligations, performing security feature reviews, employing external penetration testers to identify problems, establishing or interfacing with incident response capabilities, integrating and delivering security features, utilizing automated tools, ensuring QA performs edge/boundary value condition testing, and translating compliance constraints into requirements. These practices are distributed among the security practices outlined in the BSIMM framework.

In this paragraph, we have discussed secure software development practices within industry-created methodologies. Next, we will explore methodologies published by government entities.

\subsection{Government methodologies}\label{sec:government}

\subsubsection{The Grip on SSD methodology}\label{gripon}
The center for Information Security and Privacy Protection (CIP) was founded by the Dutch Tax Authorities to ensure the security of Dutch public services. In 2014, approximately twenty organizations formed the ``SSD practitioner community'' to share their knowledge and experience in secure software development. Since then, the community has been working on enhancing best practices in secure software development. In 2015, CIP published ``Grip on Secure Software Development (SSD)''~\cite{griponssd2015}, and that same year, 23 organizations signed the manifesto in support of the methodology. 

During our investigation of the methodology, we discovered a significant similarity between Grip on SSD~\cite{griponssd2015} and McGraw Touchpoints~\cite{mcgraw2006software}. The McGraw \textit{touchpoints} pillar is similar to the \textit{contact moments} pillar in Grip on SSD, and consists of security practices distributed among the SDLC stages. The Grip on SSD methodology also includes the \textit{standard security requirement} pillar, which outlines a set of policies, principles, and attack patterns applicable to all projects within the organization. Additionally, the McGraw's \textit{knowledge} pillar covers similar tasks, along with education and awareness initiatives.

Grip on SSD emphasizes the importance of \textit{processes} that provide guidance to clients on effectively implementing security measures. Active client support and propagation of the methodology are crucial for its success. The guidance encompasses aspects such as \textit{maturity} to determine an organization's level of control over deploying secure software, \textit{risk control and risk acceptance}, \textit{risk analysis}, \textit{business impact analysis}, and \textit{maintaining standard security requirements}.

One of the notable aspects of the Grip on SSD methodology is that the design phase primarily involves test plans based on misuse and abuse cases.

\subsubsection{NIST 800-160}

In 2022, National Institute of Standards and Technology (NIST) published ``Engineering Trustworthy Secure Systems''~\cite{NIST2016v1r1}. NIST 800-160 uses categorization of processes of ISO/IEC/IEEE 15288~\cite{iso15288}. While initially developed for engineering various systems like cyber-physical systems, Internet of Things, and hardware security, NIST 800-160 can also be applied to software engineering. 


NIST 800-160 focuses on organizational practices that span the entire SDLC of a project. \textit{Business or mission analysis} is implemented in close cooperation with the stakeholders' needs to define the drivers, scope of business and mission problems, and opportunities for problem mitigation.

The \textit{technical management} process combines risk management, decision management, configuration management, information management, and measurement. These practices collectively evaluate progress, establish and execute plans, and control project execution.  

Another significant category of security-related processes that applies to all projects is organizational \textit{project-enabling processes}. This category encompasses life cycle model management, infrastructure management, portfolio management, human resource management, quality management, and knowledge management. These practices help to ensure the organization's capabilities to fulfill project requirements.

In 2021, NIST Published Volume 2 ``Developing Cyber Resilient Systems: A Systems Security Engineering Approach''~\cite{NIST2021volume2}. This volume specifically focuses on the characteristic of cyber resilience, which is a property of engineered systems, and provides guidance on implementing cyber resilience concepts in system security engineering.

\subsubsection{The methodology from Singapore}

In 2017 Cyber Security Agency of Singapore published a white paper ``Security-by-Design Framework''~\cite{csa2017}. We will refer to this as to the methodology from Singapore. 

The methodology associates each activity with specific roles, responsibilities, and inter-dependencies. Inter-dependencies demonstrate the integration of multiple methodologies to enhance the system's security. For example, the results of a security review are utilized not only to improve security controls but also to establish the effectiveness of such controls. 

According to the authors, a key role in the security-by-design framework is the steering committee, responsible for approving milestones prior to advancing to the next phase. These milestones include security planning and risk assessment, critical security design review, system security acceptance testing, and penetration testing. 

The methodology from Singapore promotes the implementation of security-by-design principles within the Agile methodology. While security practices remain consistent with those of the Waterfall methodology, Agile introduces quick iterations in software development stages. In Agile, there is a feedback loop between the construction and the transition phases, enabling iterative and incremental delivery of stakeholder requirements. The secure practices in the construction and transition phases are similar for Agile and Waterfall methodologies, encompassing application security testing and system security acceptance testing. However, in the transition phase, Agile adds penetration testing to the mix.

\subsubsection{The methodology from Malaysia}

In 2020, CyberSecurity Malaysia, a ``national cyber security specialist agency under the purview of the Ministry of Communications and Multimedia Malaysia'', published ``Guidelines for Secure Software Development Life Cycle (SSDLC)''~\cite{malaysia2015}. We will refer to this as the methodology from Malaysia. 

During the requirement stage, the methodology incorporates \textit{data classification} to determine the protection needs for data. The data classification process involves (1) defining the type of data, (2) defining the level of sensitivity, (3) establishing data ownership, (4) implementing policy-based data management, and (5) considering privacy requirements.

In the implementation phase, the \textit{certification and accreditation} practice is used for technical verification. During deployment, the \textit{installation} practice ensures a secure production environment by encompassing activities such as environment configuration and release management. For handling change requests, the methodology employs \textit{change management}. Additionally, \textit{verification and validation} are performed during the release and maintenance phase. While the testing phase focuses on ensuring the code developed runs as intended, the main objective of \textit{verification and validation} is to confirm that the software meets the security requirements. To address residual risks during the disposal phase, the methodology includes \textit{end-of-life} policies. Organizations should adhere to these policies to ensure the proper disposal of data, documents, and software. 

\subsubsection{NIST 800-218}

In 2022, NIST released ``Secure Software Development Framework (SSDF) Version 1.1: Recommendations for Mitigating the Risk of Software Vulnerabilities''~\cite{nist218}. The framework emphasizes the preparation stage to ensure that people, technologies and processes are ready to integrate security throughout the SDLC. If needed, organization should establish new roles and responsibilities, and provide personnel training. Also, \textit{maintaining the security environment}, \textit{defining criteria for security checks}, and \textit{implementation of the supporting toolchains} are included as part of the organization's preparation.

It is important to note that the \textit{define security requirements} process in NIST 800-218 does not pertain to project-specific security requirements. Instead, it involves defining the organization's policies, risk management strategy, business objectives, applicable regulations, and more. Moreover, it is crucial to uphold the security requirements defined in these policies throughout the entire SDLC. 

Another notable practice in NIST 800-218 is \textit{design software to meet security requirements}. This practice encompasses identifying and evaluating security requirements, defining security risks, and making design decisions to mitigate these risks, thus covering both the requirements and design stages.

Upon examining NIST 800-160 and NIST 800-218, we observed that these methodologies are organization-oriented frameworks. They prioritize preparing the organization to ensure it possesses the necessary capabilities (people, processes, and technology) to undertake software development projects. To achieve this, NIST 800-218 employs the \textit{prepare the organization} category of practices, while NIST 800-160 includes \textit{organizational project-enabling processes}.

\subsubsection{Romanian SSDLC methodology}

In 2021, National Cyber Security Directorate of Romania published a paper ``Security in SDLC – Secure Software Development Lifecycle~\cite{isaca}, which we will refer to as the SSDLC methodology. This methodology shares similarities with Microsoft~\cite{microsoftsdl}, SAFECode~\cite{safecode2018}, and Malaysia~\cite{malaysia2015} methodology, as it does not introduce any distinct security practices.

In this section, we have focused on secure software development practices within government-created methodologies. Next, the methodologies published by scientific researchers are discussed.

\subsection{Academia methodologies}\label{sec:academia}

\subsubsection{The methodology by Jones and Rastogi}

In 2004, Jones and Rastogi~\cite{jones2004secure} published a software development methodology with baked-in security, which we will refer to as the methodology by Jones and Rastogi. To the best of our knowledge, this paper is the first academic publication that comprehensively describes the security practices involved in each phase of the SDLC. 

The authors mention that the methodology is based on existing risk management. However, they do not provide an explicit definition of risk management in the context of security, nor do they specify which software development stages are covered by risk management. Nevertheless, the authors provide a detailed examination of the integration between security practices and life cycle stages. 

For example, in the \textit{secure coding} practice, the authors summarize various secure coding practices, such as always authenticating, failing securely, and applying the principle of least privilege~\cite{jones2004secure}.

Another notable security practice in the methodology is the development of \textit{security test plans} during the implementation phase. These test plans are based on the system risks identified during the risk assessment practice.

The maintenance phase in this methodology encompasses organization-wide practices for  preparing the maintenance phase. This includes defining the response process for handling security bugs, establishing backup procedures, and implementing business continuity procedures. These procedures are to be prepared before entering the maintenance phase. Additionally, we consider ongoing security training for project managers, software architects, and software developers, to be an organization-wide practice. 

The security disposal phase represents the final stage of the methodology. Jones and Rastogi emphasize that security is equally important during system disposal and in the event of a disaster as it is during all other stages of the SDLC~\cite{jones2004secure}.

The authors underscore their philosophy regarding secure SDLC processes. The main idea is that organizations should provide top-level support (CEO, CFO, CIO), training for  project members involved, and management techniques that are sufficient to incorporate and support security practices. 

\subsubsection{The methodology by Apvrille and Pourzandi}

In 2005, Apvrille and Pourzandi published ``Secure Software Development by Example''~\cite{apvrille2005secure}, which we will refer to as the methodology by Apvrille and Pourzandi. The authors demonstrate secure practices through the example of PICO (Presence and Instant Communication) application, which described as a``simplified representation of ICQ or America Online Instant Messenger''~\cite{apvrille2005secure}. 

The authors' experiment revealed that code review is the most effective approach for providing security testing is a code review. Furthermore, the authors found that UMLsec is the optimal tool for illustrating security concepts during the design stage.  

\subsubsection{SSDM}

In 2006, Sodiya, Onashoga, and Ajayi published ``Towards Building Secure Software Systems'' (SSDM)~\cite{sodiya2006towards}. This framework effectively integrates the software development path with the security engineering path, while also presenting fundamental laws for creating secure software. The key laws highlighted by the authors are as follows: 
\begin{itemize}
    \item continuously update security knowledge;
    \item the software developers are required to assess their work at the end of each stage;
    \item all security specifications must be concise and clear to facilitate the implementation of security practices.
\end{itemize}

One of the notable practices within this methodology is \textit{security specification} during the design phase. The output of this practice is a policy that provides guidelines on how to effectively address attacks.

In the implementation and maintenance phase, the framework incorporates the \textit{training the users} practice. However, the authors do not provide security concerns within this particular practice. 

Furthermore, the authors utilize the \textit{implementation and maintenance phase} practice to refer to release and maintenance processes, which include software installation and implementing changes.

\subsubsection{ISDF}

In 2010, Alkussayer and Allen introduced the Integrated Security Development Framework (ISDF) Framework~\cite{alkussayer2010isdf}. This framework consists of two main elements: security best practices and the security pattern utilization process. Notable patterns include the identification of security patterns during the requirement stage, architecture evaluation during the design phase, and feedback on new patterns during the verification and operation phase. 

Furthermore, the authors emphasize that critical security concerns arise during the post-implementation phase, which falls between the deployment and operation phases. The problem lies in ensuring integrity and authenticity throughout the supply chain. To address this issue, the authors provide models for security feedback, response execution, and threat model updates. 

However, despite the authors proposing a framework that includes a visual representation of security practices involved in every SDLC phase, there is a lack of description regarding these practices in the text. For example, test plans, external reviews, and quality gates are mentioned in the framework's diagram, but the accompanying text does not provide sufficient elaboration on these aspects. As a result, we suggest that the authors have presented an inconsistent portrayal of the framework.

\subsubsection{The methodology by Daud}
In 2010, Daud published ``Secure Software Development Model: A Guide for Secure Software Life Cycle''~\cite{daud2010secure}, which we will refer to as the methodology by Daud. While the author claims that the model is based on the concept of Extreme Programming (XP), there is no mention of keywords associated with Agile methodology. For instance, the author does not utilize one-time, bucket, and every-sprint requirements, as used in MS SDL-Agile~\cite{whitepapermicrosoft} and GE methodologies~\cite{ge}. As a result, the methodology by Daud falls under the category of Waterfall methodologies in Table \ref{tab:waterfall}. Although the author presented an iterative model of secure SDLC based on the XP technique, all of these secure practices are presented within the context of the Waterfall SDLC. 

The structure of the methodology is as follows. During cycle (1), uncertain requirements are refined into well-defined requirements. Then, during cycle (2), threats identified in the analysis phase are processed in the design phase and transformed into new security requirements. Subsequently, the known security vulnerabilities and mitigation plans from the design phase are transferred to the development phase. Throughout cycle (3), implementation risks identified during development, along with user stories, are carried out over to the testing phase. The outcome of the testing phase is the identification of vulnerabilities, which are then addressed in the subsequent development phase. If design bugs are identified during testing, they should be communicated to the design phase for resolution. 

In general, the secure practices involved in Daud's Waterfall model are similar to those found in other methodologies (Jones and Rastogi~\cite{jones2004secure}, SSDM~\cite{sodiya2006towards}, ISDF~\cite{alkussayer2010isdf}). In addition, the author does not provide distinct features of his methodology. 

\subsubsection{The methodology by Khan}

In 2011, Khan published a prescriptive framework~\cite{khan2011secure} for secure software development, which we will refer to as the methodology by Khan. This methodology aims to incorporate security throughout the development lifecycle. However, the author does not explicitly highlight any unique or novel activities added to the framework. This framework appears to resemble other methodologies, such as Microsoft SDL~\cite{microsoftsdl} and Singapore~\cite{csa2017}.

\subsubsection{The methodology by Chatterje, Gupta, and De}

In 2013, Chatterjee, Gupta, and De published a framework for the development of secure software~\cite{chatterjee2013framework}, which we will refer to as the methodology by Chatterje, Gupta, and De. According to the authors, a notable feature of this methodology is the conversion of security requirements and threats into design decisions to mitigate identified security threats. However, it is worth noting that this process is a common objective in security design and is utilized in various methodologies.

This methodology shares similarities with other existing frameworks such as Microsoft SDL~\cite{microsoftsdl} and CLASP~\cite{web:lang:stats}. The methodology does not introduce any unique security practices, as evident from the information presented in Table~\ref{tab:waterfall}.

\subsubsection{The methodology by Farhan and Mostafa}

In 2018, Farhan and Mostafa published ``A Methodology for Enhancing Software Security During Development Processes''~\cite{farhan2018methodology}, where the focus is on reducing software vulnerabilities. We will refer to this methodology as the Farhan and Mostafa methodology.

According to the authors, their approach to enhancing security involves implementing security measures throughout every process and step of the SDLC, rather than solely measuring security during the testing stage. These measures are considered in RQ3. The specific security practices incorporated in the methodology are not described by the authors but are visually presented in a picture.

\subsection{Agile methodologies}
\subsubsection{SDL-Agile}\label{sdlagile}

In 2012, Microsoft released the SDL-Agile methodology~\cite{whitepapermicrosoft}, which addresses the challenge of integrating classic SDL and SDL-Agile. The main issue lies in the fact that it is not feasible to complete every requirement within a single sprint due to the limited time frame (usually 15-60 days). To address this, the authors propose two key changes to adopt classic SDL for Agile development. The first change involves reorganizing the development phases of classic SDL to better align with Agile-friendly pattern. The second change emphasized that team members should allocate sufficient SDL work to each feature before moving on to another feature. 

In addition, Microsoft identifies two main points to consider when adopting the SDL-Agile methodology: SDL-Agile requirements and the application of classic SDL tasks within sprints. For SDL-Agile requirements, the authors categorize them into three groups: every-sprint requirements, bucket requirements, and one-time requirements. These categories reflect the frequency at which the requirements need to be addressed.

Regarding SDL tasks, the authors recommend integrating threat modeling as a part of the design process in every sprint. They suggest using a ``spike'', which is a mini security push to address security issues in a specific area of code, allowing for quick updates to risky code. In the SDL-Agile project, team members can also request  exceptions for requirements during a sprint duration or for a specific period. However, in the classic SDL, exceptions are typically provided only for the entire life cycle. Moreover, the authors suggest conducting \textit{final security review} at the end of each sprint, similar to the practice in classic SDL.

\subsubsection{GE}

American General Electric Company (GE) published a white paper ``GE Digital Platform \& Product Cybersecurity (GED P\&P Cybersecurity) Secure Development Lifecycle (SDL)''~\cite{ge}, which provides guidelines for ensuring product security and reliability in Agile development. For short, we refer to this approach as the GE methodology. 

GE has proposed a framework specifically tailored for the Agile methodology, with a focus on the Industrial Internet. All the practices in this framework are categorized into three types: one-time practices, every-sprint practices, and bucket practices. In addition, GE has distributed these practices throughout the SDLC, as presented in Table~\ref{tab:agile}. 

One noteworthy practice within the GE methodology is the \textit{developer security training}, which involves ongoing courses provided to developers. However, it is worth mentioning that although this practice is continuous, the authors have categorized it as a bucket practice, which is inconsistent with the SDL-Agile approach. 

\section{Auxiliary Practices}\label{sec:nontechnical}

In this section, we answer RQ2. While investigating the secure SDLC methodologies, we discovered that they involve organizational, behavioral, legal, policy, and governance aspects, aside from purely technical aspects focused on developing software systems. The combination of these aspects we call auxiliary (non-technical practices).

\subsection{Relevance of auxiliary practices}

The influence of cultural, organizational, and personal factors on secure development has been demonstrated by researchers. For example, Arizon-Peretz, Hadar, and Luria~\cite{arizon2021importance} examined the factors affecting the implementation of the security by design approach and found that developers often lack motivation and responsibility for proactive security design due to a low level security climate and self-efficacy. The authors suggest that improving the organizational security climate could enhance the developers' self-efficacy regarding security and proactive security behavior. 

According to the 2022 Data Breach Investigation Report~\cite{dbir}, 82\% of data breaches involve human factors, which highlights the significant role played by individuals in incidents and breaches. Spiekermann, Korunovska and Langheinrich~\cite{spiekermann2018inside} conducted an experiment with 124 engineers and discovered that one-third of them did not feel motivated or responsible for designing security mechanisms. Alavi, Islam, and Mouratidis~\cite{alavi2014conceptual} argue that human factors can greatly impact security management in the organizational context, even with the presence of security measures. They identify human factors related to communication, security awareness, and management support as crucial elements.

Pirzadeh~\cite{pirzadeh2010human} revealed that human factors are often overlooked in the late phases of the software development process, despite the fact that these stages involve process improvements and maintenance based on customer satisfaction and feedback. Further research is needed to explore human factors in the late stages of SDLC and contribute to the enhancement of software development projects~\cite{pirzadeh2010human}.

Mokhberi and Beznosov~\cite{mokhberi2021sok} identified 17 factors that challenge secure software development and lead to vulnerabilities. These factors can be categorized as human, organizational, and technological. Challenges include low and high confidence level among developers, insufficient security knowledge, difficulty to grasp security concepts, lack of security culture and clear security policy, ineffective communication, misuse of security APIs/libraries and protocols, and fear to update and upgrade. To address these challenges, the authors recommend encouraging developers to utilize security knowledge and fostering a sense of responsibility, establishing security policies and strategies to support developers, promoting communication between developers and security experts, and motivating developers to enhance their security knowledge.

These studies mentioned above highlight the need for organizations to consider practices beyond traditional software development activities when adopting security-by-design approaches. Therefore, it is important to identify auxiliary practices within SSDMs.

To illustrate auxiliary practices, we categorize them into distinct groups. However, it is important to note that these categories are interconnected and often overlap, meaning that practices from one category may also apply to another. For example, practices within the \textit{understanding human behavior} category can also be relevant to the \textit{communication process} category.

\subsection{Risk management framework}

The risk management framework is defined in various methodologies, including Touchpoints~\cite{mcgraw2006software}, NIST 800-160~\cite{NIST2021volume2}, the methodology from Singapore~\cite{csa2017}, the methodology from Malaysia~\cite{malaysia2015}, the BSA~\cite{bsa2020} framework and the methodology by Jones and Rastogi~\cite{jones2004secure}. NIST 800-160~\cite{NIST2016v1r1} provides a detailed explanation of the practices involved in risk management processes. The following security activities and tasks are typically included:
\begin{itemize}
    \item \textit{planning of security risk management}. This involves defining the security aspects of the risk management strategy, taking into account stakeholders' concerns, trustworthiness, and assurance;
    \item \textit{managing the security aspects}. The information management process is involved to provide security risks to stakeholders;
    \item \textit{analysis of security risks}. With the support of the system analysis process, the analysis identifies security risks and assesses the likelihood of occurrence and consequences of these risks;
    \item \textit{treatment of security risks}. With the support of the decision management process, security treatments that may be recommended to stakeholders. 
    \item \textit{Monitoring of security risks}, which involves monitoring the changes, and assessment of the effectiveness of security measures. 
    \end{itemize}

McGraw~\cite{mcgraw2006software} describes his philosophy of the risk management framework as a full life cycle activity that occurs in parallel with SDLC activities to identify, track and mitigate risks that arise during project development. In his methodology, software risk management is strongly influenced by business motivation and takes place within the context of the business. Business goals and priorities are taken into account when identifying and analyzing risks. The risk management framework can be considered a fractal, continuous multilevel loop because the full process can be applied at different levels, such as project level, software lifecycle phase level, and artifact level.

In the methodology from Singapore~\cite{csa2017}, the BSA~\cite{bsa2020} framework, and the methodology by Jones and Rastogi~\cite{jones2004secure} the authors only mention that their methodologies are based on the risk management framework itself. However, they do not provide specific details about the framework itself.

\subsection{Security metrics}\label{secmetric}

Security metrics are measurements used to assess the effectiveness of security processes. The methodologies that mention security metrics in the secure SDLC ~\cite{mcgraw2006software,microsoftsdl,NIST2016v1r1,safecode2018,adkins2020building,web:lang:stats} highlight that there is no perfect answer to how to measure security.
 
McGraw~\cite{mcgraw2006software} considers metrics and measures to be a crucial part of introducing SDLC in large organizations. According to the author, ideally, the metrics and measures should focus on the following areas: project, process, product, and organization. By taking these areas into account, it is possible to assess all activities in a software development effort.  Moreover, all metrics should reflect strategic business goals.

NIST 800-160~\cite{NIST2016v1r1} includes the \textit{measurement process} as part of the technical management process. The main goal of the measurement process is to support effective management and demonstrate the quality of the product. The methodology also has a \textit{project assessment strategy} that addresses the measurement of security by establishing criteria for security assessment performance, methods, and evaluation activities.

Microsoft SDL~\cite{whitepapermicrosoft} and NIST 800-218~\cite{nist218} use vulnerability severity scores to define the severity threshold of security vulnerabilities and to determine the minimum acceptable security performance levels.

In contrast to the above-mentioned methodologies, CLASP~\cite{web:lang:stats} considers the role of metrics not only in assessing the likely level of security but also in identifying specific areas for improvement. CLASP metrics help assess the quality of work performed by project members. For example, the metrics can assist in deciding which part of the project requires expert attention or which project members need additional training. In CLASP, a project manager is responsible for monitoring security metrics to assess the progress of the project or the team working on a project. Compared to other methodologies~\cite{microsoftsdl, NIST2016v1r1,safecode2018,adkins2020building}, CLASP provides an overview of the metrics that can be used to measure security. These metrics include:
\begin{itemize}
    \item \textit{worksheet-based metrics}, which are based on questions regarding system assessment. Questions can be divided into the \textit{critical}, \textit{important} and \textit{useful} groups, and the metric may be based on these groups;
    \item \textit {attack surface measurement}, which is ``a count of the numbers of data inputs to the program or system''~\cite{web:lang:stats};
    \item \textit{coding guideline adherence measurement}, which allows weighting guidelines based on organizational risks;
    \item \textit{reported defect rates}, which measure the number of defects based on their severity;
    \item \textit{input validation thoroughness measurement}, which assesses whether all data from untrusted sources undergo input validation;
    \item \textit{security test coverage measurement}, which assesses the quality of testing.
\end{itemize}
In addition, the authors of CLASP highlight that it is insufficient to only identify metrics and apply them. It is crucial to consider historical metrics data and continuously track the developers' progress. The output of metrics should also be periodically reviewed.

BSIMM~\cite{bsimm12} and OWASP SAMM~\cite{owaspsamm} incorporate the \textit{strategy \& metrics} domain into their framework structure. However, it should be noted that in BSIMM~\cite{bsimm12}, none of the security activities from the top 10 activities list specifically refer to security metrics. On the other hand, OWASP SAMM~\cite{owaspsamm} does consider the definition of different metrics within the security activities in maturity levels, although it is not directly addressed in the framework itself.

The methodology by Jones and Rastogi~\cite{jones2004secure} mentions a process of establishing internal metrics and key performance indicators. However, the authors do not provide specific details about these metrics.

In the methodology by Farhan and Mostafa~\cite{farhan2018methodology}, metrics are suggested to measure security efforts in all phases of the SDLC. These metrics include: 
\begin{itemize}
    \item effort and progress metric, which measures the actual and estimated efforts and progress made;
    \item time to deliver variance rate, indicating the variance of actual progress from the baseline for the entire project;
    \item schedule variance, measuring the actual duration of the project;
    \item stability metric, illustrating the impact of requirements changes;
    \item quality measure, providing insights into quality and compliance;
    \item work product quality and software quality.
    
\end{itemize}

While the above-mentioned methodologies provide guidance on measuring security, the SAP methodology~\cite{sap2020} goes beyond that and includes the assessment of privacy. The methodology employs \textit{a data protection compliance evaluation} to assess the fulfillment of legal requirements, such as GDPR (General Data Protection Regulation). This evaluation ensures that data protection measures are aligned with applicable privacy regulations alongside security considerations in the SDLC.

\subsection{Building a culture of security}

The authors of the methodology by Google~\cite{adkins2020building} aimed to investigate the efforts to create \textit{a culture of security and reliability} within organizations, also known as a \textit{security-centric culture}. Since human factors play a crucial role in shaping security practices, it is important that everyone in the organization takes responsibility for security. 

McGraw~\cite{mcgraw2006software} described the cultural changes required to adopt SSDM in large organizations. According to his view, organizations should have a well-defined roadmap for incorporating security practices into the SDLC. This roadmap includes the following practices: (1) assigning a leader for each security initiative (2) providing training not only for developers but also for all project staff, and (3) establishing metric programs and others.

SAFECode~\cite{safecode2018} suggested that the organization's culture should be taken into account when introducing new security practices. Some organizations respond better to corporate mandates from senior managers, while others respond better to support from a team of engineers. If the organization responds better to mandates, it is advisable to designate key managers who can effectively communicate and support security initiatives.

One of the components of security culture is \textit{education and awareness} programs. These programs include appropriate training of personnel involved in the project on security basics and trends. Measuring performance outcomes also helps identify areas for improvement~\cite{nist218}. The education and awareness component has been part of the secure SDLC concept since the emergence of SSDMs. 

Over half of the methodologies we found involve ongoing security training for team members: (Microsoft SDL~\cite{microsoftsdl}, Touchpoints~\cite{mcgraw2006software}, CLASP~\cite{web:lang:stats}, NIST 800-160~\cite{NIST2016v1r1}, SAFECode~\cite{safecode2018}, the methodology by Google~\cite{adkins2020building}, BSA~\cite{bsa2020}, SAP~\cite{sap2020}, Cisco~\cite{cisco2021}, Citrix~\cite{citrix2021}, BSIMM~\cite{bsimm12}, SAMM~\cite{owaspsamm}, SDL-Agile~\cite{whitepapermicrosoft}, GE~\cite{ge}, the methodology by Jones and Rastogi~\cite{jones2004secure}, SSDM~\cite{sodiya2006towards}, ISDF~\cite{alkussayer2010isdf}, the methodology by Farhan and Mostafa~\cite{farhan2018methodology}). 

The following methodologies also address aspects of \textit{education and awareness} programs. NIST 800-160~\cite{NIST2016v1r1}includes \textit{human resource management}, which involves establishing a plan for skill development and maintaining the competence of human resources. CLASP~\cite{web:lang:stats} suggests that designating a security officer who is enthusiastic about security is a good way to increase security awareness. Furthermore, rewarding personnel for compliance with security guidelines is an effective way to raise awareness~\cite{web:lang:stats}. The authors of  SAFECode~\cite{safecode2018} claim that for successful implementation of secure SDLC, all project members need to be aware of the significance of security and attend training programs. Additionally, organizations should consider the required level of expertise for each secure practice. 

McGraw~\cite{mcgraw2006software} considers knowledge as one of the pillars. According to him, knowledge ``involves the collection, encapsulation, and sharing of security knowledge that can be used to build a solid foundation for software security practices''~\cite{mcgraw2006software}. McGraw further defines various knowledge categories, including principles, guidelines, rules, vulnerabilities, exploits, attack patterns, and historical risks. These categories are applicable throughout the software SDLC. For example, rules are utilized in static analysis and code review, while historical risks are applied to the requirement, design, implementation, and verification phases. The author argues that one of the most effective ways to disseminate software development knowledge is through security training for software development staff. 

The objective of \textit{knowledge management} in NIST 800-160~\cite{NIST2016v1r1} aligns with the knowledge pillar of Touchpoints~\cite{mcgraw2006software} and the intelligence domain of BSIMM~\cite{bsimm12}. The concept is to define, acquire, and maintain security knowledge and skills.

\subsection{Understanding human behavior}

The authors of the methodology by Google~\cite{adkins2020building} suggest that team members may sometimes experience fear or resistance to change. A successful case-building process should involve prioritizing initiatives that have a chance of success. Additionally, it is sometimes better to halt the introduction of a change if it causes more harm than benefit.

Furthermore, the authors~\cite{adkins2020building} emphasize the importance of \textit{understanding adversaries} to build secure and reliable systems. For instance, attackers may be motivated by factors such as enjoyment, recognition, activism, financial gain, coercion, manipulation, espionage, and destruction. The CLASP~\cite{web:lang:stats} methodology also analyzes the attack profile, identifying (1) insiders, (2) ``script kiddies'', (3) competitors, (4) governments, (4) organized crime, and (5) activists. In comparison, the methodology by Google includes two additional attacker profiles: automation and artificial intelligence, and vulnerability researchers. The SSDM methodology~\cite{sodiya2006towards} also emphasizes the importance of \textit{understanding attackers’ interests in the software being developed}, specifically in the security training process.

User behavior also plays a role in mitigating DoS attacks~\cite{adkins2020building}. External events and human decisions can lead to the synchronization of human behavior. For instance, during an emergency in a large city, many people may search the incident details, share information, and communicate on social networks.

The authors of the methodology by Google~\cite{adkins2020building} mention that human-centered software expertise helps address problems that users encounter while interacting with the software. Since users are not expected to have security expertise, the security of the software should not rely solely on them.

The BSA methodology~\cite{bsa2020} suggests that software should be configured securely based on its intended users' usage.

\subsection{Policies, strategies, standards and conventions}

According to NIST 800-160~\cite{NIST2016v1r1} and NIST 800-53~\cite{nist80053}, a security policy is  defined as ``a set of rules that governs all aspects of the security-relevant system and system component behavior''. The security policies and strategies establish rules and procedures for managing the security within a company. Table \ref{tab:policies} provides an overview of the security policies and strategies commonly used in SSDMs.

\begin{table*}[htbp]
  \centering
  \caption{Policies and strategies for secure software development}
  \label{tab:policies}
  \rowcolors{2}{gray!25}{white}
    \begin{tabulary}{0.8\textwidth}{p{3cm}LL}
      \rowcolor{gray!50}
    Name of policy & Meaning & Source \\
    Global security policy & Provides default standard security requirements applicable to all projects within a company & CLASP~\cite{web:lang:stats}, Grip on SSD~\cite{griponssd2015}, NIST 800-218~\cite{nist218}, the methodology by Google~\cite{adkins2020building} \\
    End-of-life (disposal) policy & Is used in managing the risks in terminating the system & Malaysia~\cite{malaysia2015}, NIST 800-160~\cite{NIST2016v1r1} \\
    Security requirements definition strategy & Are used to define common security requirements together with stakeholders & NIST 800-160~\cite{NIST2016v1r1} \\
    Project assessment strategy & Is used to measure security & NIST 800-160~\cite{NIST2016v1r1} \\
    Project control strategy & Handles problems when the project does not meet security goals & NIST 800-160~\cite{NIST2016v1r1} \\
    Decision management strategy & Includes defining roles and responsibilities, schemes to support the decision making process & NIST 800-160~\cite{NIST2016v1r1} \\
    Risk management strategy & Defines security aspects of risk management strategy & NIST 800-160~\cite{NIST2016v1r1}  \\
    Risk treatment strategy & Considers costs, schedule, and the effectiveness of reducing security risks & NIST 800-160~\cite{NIST2016v1r1}  \\
    Configuration management strategy & Involves a variety of different activities, such as roles and responsibilities, the storage media constraints, security activities among acquirer, supplier, logistics and other activities & NIST 800-160~\cite{NIST2016v1r1}  \\
    Information management strategy & Addresses security and privacy concerns of all types of information involved in the project (for example, intellectual property) & NIST 800-160~\cite{NIST2016v1r1}  \\
    Quality assurance strategy & Helps to ensure that quality management process is effectively applied for the project & NIST 800-160~\cite{NIST2016v1r1}  \\
    Vulnerability response policy & Considers vulnerability disclosure and remediation processes, roles and responsibilities & Microsoft SDL~\cite{microsoftsdl}, SAFECode~\cite{safecode2018}, BSA~\cite{bsa2020}, NIST 800-218~\cite{nist218}, the methodology by Khan~\cite{khan2011secure}, SDL-Agile~\cite{whitepapermicrosoft} \\
    Privacy policy & Is used for privacy control validation and privacy assessment practices & Cisco~\cite{cisco2021} \\
    Disaster response strategy & Defines (1) the roles and responsibilities, (2) how the incident is reported to the incident response team, and (3) communication with external stakeholders, responders, and support teams & the methodology by Google~\cite{adkins2020building} \\
    Maintenance strategy & Defines resources, security considerations, schedules, measures to perform maintenance of the system & NIST 800-160~\cite{NIST2016v1r1} \\
    Policy for authentication and authorization decisions & Verifies the identity and access rights & the methodology by Google~\cite{adkins2020building} \\
    Risk acceptance and exception policy & Considers residual risks in the deployment phase of the SDLC & the methodology from Malaysia~\cite{malaysia2015} \\
    Process management policy & Ensures that policies and procedures are consistent & NIST 800-160~\cite{NIST2016v1r1}\\
    Quality management strategy & Is oriented toward achievement of security quality objectives & NIST 800-160~\cite{NIST2016v1r1} \\
    Security requirements definition strategy & Aims to reach an agreement with stakeholders on which common security requirements must be used. The process also includes information gathering activities, methods, and techniques that are used to acquire information from stakeholders & NIST 800-160~\cite{NIST2016v1r1} \\
    Policy to control access to data and processes & Is used in identity and access management & The BSA framework~\cite{bsa2020}\\
    Coding standards & Encompasses coding rules, guidelines, and best practices & SAFECode~\cite{safecode2018}, CLASP~\cite{web:lang:stats}, the BSA framework~\cite{bsa2020}, the methodology by Google~\cite{adkins2020building}, the Cisco methodology~\cite{cisco2021}, NIST 800-218~\cite{nist218}, the methodology by Jones and Rastogi~\cite{jones2004secure}, the methodology by Farhan~\cite{farhan2018methodology}, the methodology by Apvrille and Pourzandi~\cite{apvrille2005secure}, the methodology from Singapore~\cite{csa2017}, the methodology by Khan~\cite{khan2011secure} \\
    Design standards & Provide guidance on how security features are to be used in the software design & Singapore~\cite{csa2017}, SSDLC~\cite{isaca}, the SAP methodology~\cite{sap2020}\\
    Cryptography standards & Best practices and recommendations for using encryption & Microsoft SDL~\cite{microsoftsdl} \\
    Approved tools & Are used to support engineers to use state-of-art version of tools & Microsoft SDL~\cite{microsoftsdl} \\
    Security tools & Encompass best practices for using encryption & NIST 800-218~\cite{nist218} \\
    \end{tabulary}%
\end{table*}%

Both the Grip on SSD methodology~\cite{griponssd2015} and CLASP~\cite{web:lang:stats} include the practice to have a list of baseline (or standard) security requirements that can be used for each project within a company. \textit{The standard security requirements} in SSD, as was discussed in the Section~\ref{sec:industry}, allow to avoid drawing up all security requirements afresh for each project. They include:
\begin{itemize}
    \item security architecture: in the organization of a client, some security controls may be already implemented. Security architecture defines these controls and describes the relationship between controls.=;
    \item baseline security: defines international standards that can be used in the organizations, for example, ISO 27002:2005, ISO/IEC 27002:2013, ISO 25010;
    \item classification of systems and data: the client classifies the software into security classes (high, medium, and low);
    \item risk identification: the client generates a list of known risks and then during risk analysis, relevant to current project risks are selected.
\end{itemize}

To ensure the consistency and efficiency of standard security requirements, it is recommended to exchange information on these requirements with other (semi) public bodies, allowing for the accumulation of knowledge and best practices~\cite{griponssd2015}. According to CLASP~\cite{web:lang:stats}, baseline security requirements are identified in \textit{global security policy}.

Cisco~\cite{cisco2021} believes privacy to be a fundamental human right. The company published the privacy policy~\footnote{\url{https://www.cisco.com/c/en/us/about/trust-center/global-privacy-policy.html}} which is used for privacy control validation and privacy assessment practices.

Several methodologies state that security practices need to be performed according to security policies~\cite{chatterjee2013framework, farhan2018methodology, daud2010secure, csa2017, malaysia2015, rebit2020, apvrille2005secure,khan2011secure, alkussayer2010isdf}. However, these methodologies do not define these policies. For example, the authors of the methodology from Singapore~\cite{csa2017} claim that many practices such as penetration testing, the evaluation of security specifications should be performed according to security policies. 

When writing the code, the developers may make mistakes. Defining the standards and conventions such as coding standards, languages, frameworks, and libraries helps to reduce the number of unintentional vulnerabilities in code~\cite{web:lang:stats}. SAFECode~\cite{safecode2018}, CLASP~\cite{web:lang:stats}, and BSA~\cite{bsa2020} involve \textit{establish coding standards} as a part of secure coding practices. While CLASP does not explicitly designate coding standards as a specific practice, it does provide a list of recommended coding standards. The methodology by Google~\cite{adkins2020building}, Cisco~\cite{cisco2021}, NIST 800-218~\cite{nist218}, the methodology by Jones and Rastogi~\cite{jones2004secure}, the methodology by Farhan and Mostafa~\cite{farhan2018methodology}, the methodology by Apvrille and Pourzandi~\cite{apvrille2005secure} include coding standards, but identification of these standards is not the practice of the methodologies.

Both BSIMM~\cite{bsimm12} and OWASP SAMM~\cite{owaspsamm} include the governance domain, which helps organize, manage and measure security activities. They incorporate practices such as \textit{strategies \& metrics} and \textit{compliance \& policy} practices, \textit{education and guidance} and \textit{training}. These methodologies also include activities related to establishing the security, which are grouped into maturity levels. As the security activities involved in maturity levels lie beyond the scope of the research, we do not include specific security policies in Table~\ref{tab:policies}.

\subsection{Auxiliary practices of incident or vulnerability response}

In the realm of software development, where humans are prone to making mistakes~\cite{adkins2020building,isaca}, it becomes crucial to detect and address these issues early in the SDLC to minimize the costs associated with rectifying such mistakes~\cite{howard2006security}. However, in released software mistakes and vulnerabilities may still exist. To tackle this challenge, various methodologies, such as Microsoft SDL~\cite{microsoftsdl}, SAFECode~\cite{safecode2018}, CLASP~\cite{web:lang:stats}, NIST 800-218~\cite{nist218}, the methodology by Google~\cite{adkins2020building}, Cisco~\cite{cisco2021}, and Citrix~\cite{citrix2021} employ auxiliary practices for incident or vulnerability response.

Microsoft SDL~\cite{microsoftsdl} advocates for the establishment of a security response center within organizations. This center comprises individuals responsible for responding to externally discovered vulnerabilities and collaborating with security researchers who have uncovered these vulnerabilities. The response center maintains communication with the researchers, providing them with updates on the status of the response and update process. Building relationships of confidence and trust with vulnerability researchers is emphasized, as it helps to reduce customer exposure to vulnerabilities until a fix is released. Techniques such as simple and personal communication, recognizing the value of vulnerability researchers, conducting partner programs to provide early access for researchers, and granting them access to test updates, can all foster collaboration with vulnerability researchers.

Similarly, SAFECode~\cite{safecode2018} emphasizes the importance of maintaining contact with reporters and promptly communicating the availability of vulnerability fixes to customers. The communication processes with customers and security researchers are also described in CLASP~\cite{web:lang:stats} and NIST 800-218~\cite{nist218}. Organizations like Cisco~\cite{cisco2021} and Citrix~\cite{citrix2021} have dedicated Product Security Incident Response Team (PSIRT) to handle communication with customers.

The methodology by Google~\cite{adkins2020building} focuses on human behavior while preparing for and being in \textit{incident or disaster}. This strategy encompasses analyzing potential disasters, establishing response teams, creating response plans, configuring systems properly, testing procedures, and seeking feedback. These components of preparation for disaster are considered next.

According to the methodology by Google~\cite{adkins2020building}, the formalization of team structure, information management, and communication between the recovery team are vital components for \textit{recovering from the incident}. The scope of recovery depends on the type of the attack. After recovering the system and ejecting the attacker, organizations should consider the impacts of the attack. This will help improve incident handling. While analyzing the impact, the following questions may be useful to consider (1) what are the main factors that contributed to the incident? (2) how quickly was the incident detected? (3) how may the detection system be improved? 

To prevent total disruption of the system, organizations should conduct a disaster risk analysis. This analysis includes the following steps: (1) identifying human or technological resources required to respond to an incident, (2) identifying potential disaster scenarios that may occur in the system, and (3) identifying systems that, if disabled or disrupted, can disable operations. When developing a response plan, organizations should create high-level procedures that define (1) the roles and responsibilities, (2) how incident are reported to the incident response team, and (3) communications with external stakeholders, responders, and support teams. Organizations should  also train engineers in response activities and provide feedback to prevent the same mistakes.

\subsection{Communication process and customer responsibilities}

In SAFECode~\cite{safecode2018}, \textit{stakeholder management and communication} involves explaining to stakeholders the value and commitment to secure development practices.

\textit{The processes} pillar of Grip on SSD~\cite{griponssd2015} includes business impact analysis (BIA), which aims to establish ``quality requirements for the information systems used within that operational process''~\cite{griponssd2015}. The client is one of the responsible parties in BIA, with the following responsibilities:
\begin{itemize}
    \item determining the goals of operational processes within the organization's main tasks;
    \item identifying the main and auxiliary sub-processes;
    \item verifying the execution of the BIA process.
\end{itemize}

Another non-technical practice in Grip on SSD involving the client is risk acceptance before the release phase. Risk acceptance requires the client, supported by a security advisor, to decide on accepting risks. There are three options available to the client: (1) accepting compliant software, (2) modifying non-compliant software, and (3) temporarily allowing non-compliant software. If the client chooses temporary acceptance, the following points should be considered:
\begin{itemize}
    \item the plan outlining when and how the solution will be presented;
    \item the budget required for implementing the plan;
    \item the client's approval of the plan.
\end{itemize}

Meanwhile, in the methodology from Malaysia~\cite{malaysia2015}, the customer is responsible for accepting residual risks.

The authors of the methodology by Google~\cite{adkins2020building} emphasize that during tight deadlines and high stress, communications with team members and external parties may be challenging. Misunderstandings can arise as a result of these communication difficulties. To mitigate this problem, it is recommended to be overly communicative and explicit. Another challenge is hedging, which often introduces confusion and uncertainty into the decision-making process. Regular, well-managed meetings help maintain control and visibility of ongoing activities. Lastly, determining the appropriate level of detail to share is another communication challenge that needs to be addressed. 

\subsection{Ethics}

The topic of ethics in secure software development is addressed in CLASP~\cite{web:lang:stats}. The organizations as a whole are expected to uphold ethics standards, although it may not be realistic to expect every individual component to be inherently ethical. An important consideration is the unethical behavior of insiders who may attack the organization, and the organization should take this into account. Ethical behavior, in general, entails providing users with a privacy policy, notifying them of any changes in the policy, and promptly informing them in the event of a privacy breach.

\subsection{Privacy}

The topic of privacy is addressed in Microsoft SDL v.5.2~\cite{whitepapermicrosoft} and SDL-Agile~\cite{whitepapermicrosoft}, which both focus on privacy requirements. Within these frameworks, a privacy advisor is assigned to provide support. However, the primary responsibility for privacy lies with the privacy lead, who is a member of the project team. In SDL-Agile, reporting design changes that impact privacy to the privacy advisor is the every-sprint requirement. During the release phase in Microsoft SDL V.5.2, it is crucial to collaborate with the privacy advisor and legal representatives to create an approved privacy disclosure.

According to the methodology by Google~\cite{adkins2020building}, organizations should have the capability to investigate systems after a failure. Therefore, it is necessary for organizations to design a logging system with access control and protection. Privacy and legal members should be involved in the design process of the logging system.

Cisco~\cite{cisco2021} incorporated a \textit{privacy assessment} to evaluate privacy controls based on laws and regulations. Additionally, Cisco provides a dedicated privacy Trust Portal~\footnote{\url{https://trustportal.cisco.com/c/r/ctp/trust-portal.html}} for customers to understand the data processing procedures. 

Privacy requirements and controls are also considered in the methodology from Malaysia~\cite{malaysia2015}. For example, measures such as data anonymization, disposition, and pseudonymization are implemented.

\section{Evaluation of the methodologies}

\subsection{Evaluation of the methodologies in academic research}

One of the common approaches to assess the benefits of the methodology for an organization is through conducting case studies~\cite{391832}. In the reviewed literature, case studies were found to be the most prevalent method used to evaluate the effectiveness of proposed methodologies.

The methodology by Apvrille and Pourzandi~\cite{apvrille2005secure} was presented using an instant messaging application as an example. Although the evaluation of the methodology was beyond the scope of their research, the authors believe that the methodology can enhance the security level of the software.

The authors of the SSDM methodology~\cite{sodiya2006towards} conducted a case study by implementing it in an accounting system. During the system's 3 years usage, there were 129 security breaches. However, after implementing SSDM, no security breaches were found during one year of usage. The case study results demonstrate an improvement in security. However, it is unclear whether the company had previously employed any other SSDM before the case study.

The authors of ISDF~\cite{alkussayer2010isdf} built an e-commerce system to showcase the advantages of their methodology. However, the authors provided examples for the requirement and design stages, lacking demonstrations of the methodology's effectiveness in other activities.

Chatterjee, Gupta, and De~\cite{chatterjee2013framework} conducted a case study on a web-based banking system, focusing on security design. Similar to ISDF~\cite{alkussayer2010isdf}, the authors only involved the requirements and the design phase. Additionally, the authors compared their methodology with the methodology by Apvrille and Pourzandi~\cite{apvrille2005secure} and the AOD approach~\cite{georg2002using}, which proposes security aspects for the design stage. The results indicate that the methodology by Chatterjee, Gupta, and De suggests more suitable design decisions than the other methodologies~\cite{apvrille2005secure,georg2002using}. However, this case study does not provide conclusive evidence of the effectiveness of the methodology.

The methodology by Jones and Rastogi~\cite{jones2004secure}, the methodology by Daud~\cite{daud2010secure}, the methodology by Khan~\cite{khan2011secure}, the methodology by Farhan and Mostafa~\cite{farhan2018methodology} did not offer any experiments or evidence of their effectiveness.

\subsection{Evaluation of the methodologies from the industry}

During the investigation of industry and government SSDMs, we discovered that none of the methodologies provide evidence of effectiveness. 

The authors of NIST 800-218~\cite{nist218} argue that incorporating the security practices mentioned in the methodology can help reduce the number of vulnerabilities in software. However, they do not present experimental results or other evidence to support this claim.

CLASP~\cite{web:lang:stats}, SAFECode~\cite{safecode2018}, Grip on SSD~\cite{griponssd2015}, the methodology by Google~\cite{adkins2020building}, the SAP methodology~\cite{sap2020}, the Cisco methodology~\cite{cisco2021} and the Citrix methodology~\cite{citrix2021} emphasize that their methodologies are the results of years of experience and aim to provide a set of security best practices. SAP has a blog~\cite{saptrust} where the Head of Product Security SAP compared the SAP methodology with NIST 800-218. The comparison reveals that almost all the practices and recommendations in NIST 800-218 have corresponding measures and controls in the SAP methodology. This example demonstrates one of the ways of evaluating a methodology by comparing it with established security standards. 

Microsoft's website, specifically the Frequently Asked Questions page, contains information stating that ``The SDL has proven to be effective at reducing vulnerability counts of flagship Microsoft products after release''~\cite{microsoftsdl}. In 2023 Lipner and Howard~\cite{lipner2023inside} published the results of their security push released in 2003~\cite{howard2003inside} and an evaluation of the Secure SDL~\cite{howard2006security}. The authors assert that there has been a significant reduction in the number of vulnerabilities in Microsoft software products, which indicates the validation of the implemented security measures. They also claim that the evidence of Microsoft SDL effectiveness lies in the code, output of security tools, and threat models~\cite{lipner2023inside}.

McGraw~\cite{mcgraw2006software} ranks the touchpoints according to their effectiveness and importance. According to the author~\cite{mcgraw2006software}, the ranking is based on the experience of applying touchpoints in different organizations. However, the author does not provide concrete evidence of their effectiveness, such as  case studies or experimental results.

Both maturity models, SAMM~\cite{owaspsamm} and BSIMM~\cite{bsimm12}, allow organisations to assess the maturity of their software development process and provide an overview of the status of security activities. However, the maturity level does not reflect the effectiveness of the security process. Thus, neither SAMM nor BSIMM allow for the assessment of effectiveness of security efforts.

\subsection{Other literature on assessing effectiveness of SSDM}
In addition to the authors of the studied methodologies, numerous researchers have explored various methods for assessing the effectiveness of secure software engineering. Busch, Koch, and Wirsing~\cite{busch2014evaluation} introduced the SecEval method for evaluating engineering approaches in the SDLC. According to the authors, SecEval enables a ``structured evaluation of methods, tools, notations, security properties, vulnerabilities and threats''~\cite{busch2014evaluation}. The model consists of three components: (1) a context model that describes security properties, threats and vulnerabilities, (2) data collection model that records how data is gathered, (3) a data analysis model that specifies how reasoning is performed based on the collected data.

The Dagstuhl seminar ``Empirical
Evaluation of Secure Development Processes~\cite{shostack2019empirical} covered various important topics relevant to our research. One of the key discussions was ``How do we know that the system is really secure?''. Bodden~\cite[p.21]{shostack2019empirical} suggested that software security metrics should consider the assume-breach paradigm, which refers to the ability of software to withstand attacks despite known and unknown vulnerabilities in the system. In other words, metrics should not assume that the software is free of vulnerabilities. The author argues that establishing measurable indicators of security can facilitate the creation of effective software security metrics. Additionally, Weber et al.~\cite[p.23]{shostack2019empirical} discussed the empirical evaluation of software development processes. According to the authors, obtaining a comprehensive understanding of the advantages and disadvantages of a particular development methodology would likely require employing multiple techniques. 


\section{Discussion}\label{sec:discussion}

In this section, we discuss the key findings regarding our research questions. 

To answer RQ1, we have discovered 28 SSDMs published between 2004 and 2022. These methodologies originate from industry, governments and academic researchers. The majority of the discovered methodologies have emerged from large companies. Based on the summary of security practices shown in Table~\ref{tab:waterfall} and Table~\ref{tab:agile}, we observed that these methodologies have not undergone substantial evolution since 2004. Even in the earliest methodologies, auxiliary practices, such as organizational, behavioral, legal, policy, governance aspects, and common technical security practices were considered. However, over the span of 18 years, some methodologies, such as NIST 800-218~\cite{nist218} and BSA~\cite{bsa2020}, have become more specific by providing the references to the standards for each security practice.

During the mapping of the security practices to the SDLC stages, we identified certain auxiliary practices that encompassed cultural, organizational, and personal factors. These practices were combined to address RQ2, resulting in the identification of nine intertwined categories of auxiliary practices. These categories include risk management, security measurement, building a culture of security, understanding human behavior, creating policies and strategies, and communication processes. 

For RQ3, we investigated the methods used by the authors to provide evidence of the effectiveness of the secure software development process. We discovered that most of the methodologies imply their effectiveness and their contribution to software security improvement but do not provide concrete evidence. Out of the eight academic papers reviewed, only one methodology included a case study, while two methodologies involved case studies related to security practices in the requirements and/or design stages. Industry and government methodologies do not provide any evidence of their effectiveness.

During our investigation of the methodologies, we identified significant gaps that need to be addressed to enhance software security. For example, as previously discussed, there is a need to explore ways to assess the effectiveness of SSDMs. Another notable gap is the scarcity of auxiliary (non-technical) practices in academic papers. The only practices considered are the \textit{risk management framework} (in the methodology by Jones and Rastogi~\cite{jones2004secure}) and \textit{education and awareness} (in the methodology by Jones and Rastogi~\cite{jones2004secure}, SSDM~\cite{sodiya2006towards}, ISDF~\cite{alkussayer2010isdf}, the methodology by Farhan and Mostafa~\cite{farhan2018methodology}). As discussed in our response to RQ2 (Section~\ref{sec:nontechnical}), there are numerous other auxiliary practices, such as privacy, ethics, human behavior, and communication processes. All these auxiliary practices are derived exclusively from industry and government methodologies.

Additionally, after conducting this survey, two questions have arisen:

\begin{itemize}
    \item Why are there so many methodologies, and why do new methodologies continue to emerge?
    \item Why do data breaches still occur even when all stages of the secure SDLC are followed?
\end{itemize}

To answer these questions, further research is required.

\section{Related work}\label{sec:relwork}
\subsection{Existing literature reviews}
To the best of our knowledge, our literature review represents the most comprehensive effort to examine the existing SSDMs. However, there have been other related endeavours have explored security practices within the software development process. A summary of existing literature reviews that address at least one of our research questions can be found in Table~\ref{tab:reviews}. 

While many of these literature reviews share similarities with ours in terms of the included SSDMs, some of them delve into research questions that extend beyond the scope of our study. For instance, Williams~\cite{williamssecure} investigated the integration of SSDM with various software development models, such as Agile and DevOps, mobile applications, IoT, cloud computing, road vehicles and E-commerce. 

N{\'u}{\~n}ez, Lindo and Rodr{\'\i}guez~\cite{nunez2020preventive} proposed the Viewnext-UEx model, which incorporates security practices from established models while addressing their weaknesses. The model introduces new security practices, namely the \textit{state of the project}, \textit{security observatory} and \textit{vulnerabilities repository}. The first practice aims to evaluate projects from a security perspective, ensuring compliance with the security guidelines. The second practice focuses on reducing the time spent in an insecure state by actively searching for new attack techniques and vulnerabilities. The last practice involves building a knowledge base from security failures and errors to enhance developers' training. 

Additionally, Ramirez, Aiello and Lincke~\cite{ramirez2020survey} not only analyzed SSDMs but also considered standards and certifications, such as Common Criteria and The Open Group Architecture Framework. 

\begin{table*}[h!]
\centering
  \caption{Chronological summary of literature reviews}
  \label{tab:reviews}
    \begin{tabular}{p{2em}p{30em}ccccc}
    \hline
    \multirow{2}{*}{Year} & \multirow{2}{*}{Author and title} & \multirow{2}{*}{\begin{tabular}[c]{@{}l@{}}Count of meth.\\ investigated\end{tabular}} & \multicolumn{4}{p{13em}}{Alignment with our RQs } \\ 
       &  &  & RQ1     & RQ2     & RQ3     & RQ4 \\ \hline

    2005 & Davis ``Secure software development life cycle processes: A technology scouting report''~\cite{davis2005secure} & 10 &  \checkmark{} & $\times$ & $\times$ & $\times$ \\ 
   
    2007 & Gregoire et al. ``On the secure software development process: CLASP and SDL compared''~\cite{gregoire2007secure} & 2 & \checkmark{} & $\times$ & $\times$ & \checkmark{} \\ 

    2008     & De Win et al. ``On the secure software development process: CLASP, SDL and Touchpoints compared'' \cite{de2009secure} &  3 & \checkmark{} & $\times$ & $\times$ & \checkmark{} \\ 
    
    2013 & Fonseca and Vieira ``A survey on secure software development lifecycles''~\cite{fonseca2013survey} & 4 & \checkmark{} & $\times$ & $\times$ & $\times$ \\
    
    2019 & Williams  ``Secure software lifecycle knowledge area''~\cite{williamssecure} & 3 & \checkmark{} & $\times$ & $\times$ & $\times$ \\
    
    2020 & N{\'u}{\~n}ez, Andr{\'e}s, and Rodr{\'\i}guez ``A preventive secure software development model for a software factory: A case study''~\cite{nunez2020preventive} & 7  & \checkmark{} & \checkmark{} & $\times$ & $\times$ \\ 
    
    2020 & Ramirez, Aiello, and Lincke ``A survey and comparison of secure software development standards''~\cite{ramirez2020survey} & 24 & \checkmark{} & $\times$ & $\times$ & $\times$ \\ 

    \end{tabular}%
\end{table*}%

We applied the same inclusion and exclusion criteria as  presented in Table~\ref{tab:criteria} when selecting studies for Table~\ref{tab:reviews}. Certain studies were excluded because they only investigated security practices in specific stages of SDLC rather than spanning the entire SDLC. 

One such work is the maturity model proposed by Al-Matouq et al.~\cite{al2020maturity} for secure software design, which is based on security practices identified in a comprehensive study. Their study's scope extends beyond ours, encompassing not only SSDMs, but also maturity models and methodologies for the software design phase. Moreover, their research~\cite{al2020maturity} focuses not solely on methodologies, but also on papers that incorporate security practices. 

Khan et al.~\cite{khan2021systematic} investigated security approaches in secure software engineering. The authors concentrated on security practices within different phases of the SDLC. Additionally, Khan et al.~\cite{khan2022systematic} explored security risks and associated practices in secure software development.

\subsection{Other relevant literature}

Several researchers have investigated the interaction between developers and secure software processes. For example, Acar et al.~\cite{acar2017developers} argue that developers need human-centered security experts and legal experts to solve social engineering problems. Assal and Chiasson~\cite{assal2019think} claim that organizational issues, such as the lack of security plans, procedures, knowledge, or resources, are the primary reasons for postponing security.

There are also studies that explored security models beyond the scope of this research. For instance, Myrbakken and Colomo-Palacios~\cite{myrbakken2017devsecops} conducted a literature review on DevSecOps methodologies. Sánchez-Gordón and Colomo-Palacios~\cite{sanchez2020security} conducted a literature review to understand the DevSecOps culture from a human factor perspective. Cybersecurity capability maturity models have been investigated in studies, such as~\cite{stevanovic2011maturity,rabii2020information,le2016can,rea2017maturity}. 

Certain researchers have explored security methodologies in specific areas. Uzunov, Fernandez, Falkner~\cite{uzunov2012engineering} investigated security methodologies applicable to distributed systems. Suganya, Jothi, and Palanisamy~\cite{suganya2018survey} explored security methodologies in e-voting systems. Malik and Nazir~\cite{malik2012security} studied security frameworks for the cloud computing environment. Babar et al.~\cite{babar2011proposed} proposed an embedded security framework for the Internet of Things (IoT). Pacheco and Hariri~\cite{pacheco2016iot} developed an IoT security framework for smart infrastructures. ENISA~\cite{enisa} introduced practices for IoT security. Agile security methods are considered in~\cite{rindell2017busting}. Kang and Kim~\cite{kang2022cia} proposed a CIA (functional correctness, safety integrity, security assurance)-level framework that provides security measures to establish the required  level of security in organizations. The framework also allows for comparison of security process levels with competitors through gap analysis. Ardo, Bass and Gaber~\cite{ardo2022towards} developed a methodology based on interviews with Agile practitioners. However, as the paper meets exclusion criteria EX-4~\ref{tab:criteria}, we did not include it in our research.

Although there is an exhaustive body of literature focused on secure development activities and methodologies, several studies have identified the challenges that organizations and developers face when adopting secure software development practices. Maher et al.~\cite{maher2020challenges} revealed that a lack of clear vision, inadequate guidelines from top management, and insufficient guidance on how to incorporate security practices pose challenges to the adoption of secure software development. Gasiba et al.~\cite{gasiba2020awareness} also discovered that while developers are motivated to produce secure code, the lack of knowledge of secure coding guidelines hinders their ability to do so. 

In a study by Kirlappos, Beautement and Sasse~\cite{kirlappos2013comply}, key reasons for non-compliance with organizational policies were observed. The authors concluded that the adoption of security practices should be decentralized, allowing employees to determine how to incorporate security into their individual tasks. Additionally, a survey conducted by Geer~\cite{geer2010companies} revealed that organizations do not widely adopt formal secure SDLC framework due to challenges such as a lack of awareness of methodologies and the perceived time consuming nature of implementing these methodologies.

\section{Conclusion}\label{sec:conclusion}

In this survey, we collected 28 SSDMs from industry, government, and academia sectors. During the mapping of security practices to the SDLC phases, we observed that the SDLC process involves not only purely technical practices but also auxiliary practices. These auxiliary practices include measuring security, fostering a culture of security, developing policies and strategies, promoting effective team communication, ethics and privacy considerations.

Upon investigating how authors provide evidence of the effectiveness of their methodologies, we discovered that most of the methodologies imply their effectiveness in improving software security but fail to provide concrete evidence. In fact, some methodologies do not even mentioned effectiveness at all. Among the eight academic papers reviewed, only one methodology included a case study, while two methodologies involved case studies specifically focused on security practices in the requirements and/or design stages. 

As a result of this survey, several research gaps have been identified. One open question is why companies tend to create their own methodologies instead of adopting existing ones. Another research gap pertains to the lack of evidence supporting effectiveness of these methodologies, often based on the belief that they reduce the number of vulnerabilities in software. Authors commonly do not provide factual support for their beliefs. Additionally, academic methodologies tend to sparingly incorporate auxiliary (non-technical) security practices, with a primary focus on technical security practices. Some academic methodologies even lack information on what is novel about their approach, relying on security practices already published in existing methodologies. Lastly, despite the availability of numerous SSDMs, there is a concerning trend of increasing vulnerabilities in software. We believe that addressing these identified gaps can contribute to the development of software with fewer vulnerabilities. Exploring these gaps provides a foundation for future research in this area.

\section*{Acknowledgment}
This research has been partially supported by the Dutch Research Council (NWO) under the project NWA.1215.18.008 Cyber Security by Integrated Design (C-SIDe).




\bibliographystyle{IEEEtran}
\bibliography{bibliography.bib}

\begin{thebibliography}{100}
\providecommand{\url}[1]{#1}
\csname url@samestyle\endcsname
\providecommand{\newblock}{\relax}
\providecommand{\bibinfo}[2]{#2}
\providecommand{\BIBentrySTDinterwordspacing}{\spaceskip=0pt\relax}
\providecommand{\BIBentryALTinterwordstretchfactor}{4}
\providecommand{\BIBentryALTinterwordspacing}{\spaceskip=\fontdimen2\font plus
\BIBentryALTinterwordstretchfactor\fontdimen3\font minus
  \fontdimen4\font\relax}
\providecommand{\BIBforeignlanguage}[2]{{%
\expandafter\ifx\csname l@#1\endcsname\relax
\typeout{** WARNING: IEEEtran.bst: No hyphenation pattern has been}%
\typeout{** loaded for the language `#1'. Using the pattern for}%
\typeout{** the default language instead.}%
\else
\language=\csname l@#1\endcsname
\fi
#2}}
\providecommand{\BIBdecl}{\relax}
\BIBdecl

\bibitem{cvemitre}
{MITRE}, ``{CVE} details,'' available at
  \url{https://www.cvedetails.com/browse-by-date.php}. Accessed in May 2023.

\bibitem{forbes}
C.~Brooks, ``Alarming cybersecurity stats: What you need to know for 2021,''
  available at
  \url{https://www.forbes.com/sites/chuckbrooks/2021/03/02/alarming-cybersecurity-stats-------what-you-need-to-know-for-2021/?sh=5ff0d7a658d3}.
  Accessed in May 2023.

\bibitem{microsoftaboutsdl}
Microsoft, ``About microsoft {SDL},'' 2020, available at
  \url{https://www.microsoft.com/en-us/securityengineering/sdl/about}. Accessed
  in May 2023.

\bibitem{howard2006security}
M.~Howard and S.~Lipner, \emph{The security development lifecycle}.\hskip 1em
  plus 0.5em minus 0.4em\relax Microsoft Press Redmond, 2006, vol.~8.

\bibitem{mcgraw2004software}
G.~McGraw, ``Software security,'' \emph{IEEE Security \& Privacy}, vol.~2,
  no.~2, pp. 80--83, 2004.

\bibitem{wing2003call}
J.~M. Wing, ``A call to action look beyond the horizon,'' \emph{IEEE Security
  \& Privacy}, vol.~1, no.~6, pp. 62--67, 2003.

\bibitem{mcgraw2002building}
G.~McGraw, ``Building secure software: better than protecting bad software,''
  \emph{IEEE Software}, vol.~19, no.~6, pp. 57--58, 2002.

\bibitem{mcgraw2003ground}
------, ``From the ground up: The dimacs software security workshop,''
  \emph{IEEE Security \& Privacy}, vol.~1, no.~2, pp. 59--66, 2003.

\bibitem{leblanc2002writing}
D.~LeBlanc and M.~Howard, \emph{Writing secure code}.\hskip 1em plus 0.5em
  minus 0.4em\relax Pearson Education, 2002.

\bibitem{viega2001building}
J.~Viega and G.~R. McGraw, \emph{Building secure software: How to avoid
  security problems the right way, portable documents}.\hskip 1em plus 0.5em
  minus 0.4em\relax Pearson Education, 2001.

\bibitem{sap2020}
``The secure software development lifecycle at {SAP},'' SAP, White paper, 2020,
  available at
  \url{https://www.sap.com/documents/2016/03/a248a699-627c-0010-82c7-eda71af511fa.html}.
  Accessed in May 2023.

\bibitem{citrix2021}
``Citrix security development lifecycle,'' Citrix, White paper, 2021, available
  at
  \url{https://www.citrix.com/content/dam/citrix/en_us/documents/about/citrix-security-development-lifecycle.pdf}.
  Accessed in May 2023.

\bibitem{cisco2021}
``Cisco secure development lifecycle,'' Cisco, White paper, 2021, available at
  \url{https://www.cisco.com/c/dam/en_us/about/doing_business/trust-center/docs/cisco-secure-development-lifecycle.pdf}.
  Accessed in May 2023.

\bibitem{de2009secure}
B.~De~Win, R.~Scandariato, K.~Buyens, J.~Gr{\'e}goire, and W.~Joosen, ``On the
  secure software development process: {CLASP, SDL} and {T}ouchpoints
  compared,'' \emph{Information and software technology}, vol.~51, no.~7, pp.
  1152--1171, 2009.

\bibitem{davis2005secure}
N.~Davis, ``Secure software development life cycle processes: A technology
  scouting report,'' Carnegie-Mellon University, 2005.

\bibitem{gregoire2007secure}
J.~Gregoire, K.~Buyens, B.~De~Win, R.~Scandariato, and W.~Joosen, ``On the
  secure software development process: {CLASP} and {SDL} compared,'' in
  \emph{Proceedings of the Third International Workshop on Software Engineering
  for Secure Systems}.\hskip 1em plus 0.5em minus 0.4em\relax IEEE, 2007, pp.
  1--1.

\bibitem{fonseca2013survey}
J.~Fonseca and M.~Vieira, ``A survey on secure software development
  lifecycles,'' in \emph{Software Development Techniques for Constructive
  Information Systems Design}.\hskip 1em plus 0.5em minus 0.4em\relax IGI
  Global, 2013, pp. 57--73.

\bibitem{williamssecure}
L.~Williams, \emph{Secure Software Lifecycle Knowledge Area}, 2019, available
  at
  \url{https://www.cybok.org/media/downloads/Secure_Software_Lifecycle_issue_1.0.pdf}.
  Accessed in May 2023.

\bibitem{nunez2020preventive}
J.~C.~S. N{\'u}{\~n}ez, A.~C. Lindo, and P.~G. Rodr{\'\i}guez, ``A preventive
  secure software development model for a software factory: a case study,''
  \emph{IEEE Access}, vol.~8, pp. 77\,653--77\,665, 2020.

\bibitem{ramirez2020survey}
A.~Ramirez, A.~Aiello, and S.~J. Lincke, ``A survey and comparison of secure
  software development standards,'' in \emph{Proceedings of the 13th CMI
  Conference on Cybersecurity and Privacy (CMI)-Digital
  Transformation-Potentials and Challenges}.\hskip 1em plus 0.5em minus
  0.4em\relax IEEE, 2020, pp. 1--6.

\bibitem{pirzadeh2010human}
L.~Pirzadeh, ``Human factors in software development: a systematic literature
  review,'' \emph{Citeseer}, 2010.

\bibitem{mokhberi2021sok}
A.~Mokhberi and K.~Beznosov, ``So{K}: Human, organizational, and technological
  dimensions of developers’ challenges in engineering secure software,'' in
  \emph{Proceedings of the European Symposium on Usable Security 2021}, 2021,
  pp. 59--75.

\bibitem{garousi2019guidelines}
V.~Garousi, M.~Felderer, and M.~V. M{\"a}ntyl{\"a}, ``Guidelines for including
  grey literature and conducting multivocal literature reviews in software
  engineering,'' \emph{Information and software technology}, vol. 106, pp.
  101--121, 2019.

\bibitem{isaca}
T.~Adam, F.~Andrei, L.~Gabudeanu, and V.~Rotaru, ``Security in {SDLC} –
  secure software development lifecycle – {SSDLC},'' White paper, 2021,
  available at \url{https://dnsc.ro/vezi/document/security-in-sdlc}. Accessed
  in May 2023.

\bibitem{apvrille2005secure}
A.~Apvrille and M.~Pourzandi, ``Secure software development by example,''
  \emph{IEEE Security \& Privacy}, vol.~3, no.~4, pp. 10--17, 2005.

\bibitem{microsoftsdl}
Microsoft, ``{SDL},'' available at
  \url{https://www.microsoft.com/en-us/securityengineering/sdl/practices}.
  Accessed in May 2023.

\bibitem{ge}
``Secure development lifecycle,'' GE, White paper, available at
  \url{https://www.ge.com/digital/documentation/predix-platforms/sdl.html}.
  Accessed in May 2023.

\bibitem{nist218}
``Secure software development framework {(SSDF)},'' National Institute of
  Standards and Technology, Version 1.1, 2022, available at
  \url{https://doi.org/10.6028/NIST.SP.800-218}. Accessed in May 2023.

\bibitem{csa2017}
``Security-by-design framework,'' Cyber Security Agency of Singapore, Version
  1.0, 2017, available at
  \url{https://www.csa.gov.sg/docs/default-source/csa/documents/legislation_supplementary_references/security_by_design_framework.pdf?sfvrsn=560b9ff3_0}.
  Accessed in May 2023.

\bibitem{bsa2020}
``The {BSA} framework for secure software,'' BSA, Version 1.1, 2020, available
  at
  \url{https://www.bsa.org/files/reports/bsa_framework_secure_software_update_2020.pdf}.
  Accessed in May 2023.

\bibitem{alkussayer2010isdf}
A.~Alkussayer and W.~H. Allen, ``The {ISDF} framework: towards secure software
  development,'' \emph{Journal of Information Processing Systems}, vol.~6,
  no.~1, pp. 91--106, 2010.

\bibitem{khan2011secure}
R.~Khan, ``Secure software development: a prescriptive framework,''
  \emph{Computer Fraud \& Security}, vol. 2011, no.~8, pp. 12--20, 2011.

\bibitem{chatterjee2013framework}
K.~Chatterjee, D.~Gupta, and A.~De, ``A framework for development of secure
  software,'' \emph{{CSI Transactions on ICT}}, vol.~1, no.~2, pp. 143--157,
  2013.

\bibitem{griponssd2015}
M.~Koers, R.~Paans, R.~van~der Veer, C.~Kok, and J.~Breeman, ``Grip on secure
  software development {(SSD)},'' CIP, Version 2.0, 2015, available at
  \url{https://www.cip-overheid.nl/media/1105/20160622_grip_on_ssd_the_method_v2_0_en.pdf}.
  Accessed in May 2023.

\bibitem{malaysia2015}
``Guidelines for secure software development life cycle {(SSDLC)},'' Ministry
  of {C}ommunications and {M}ultimedia {M}alaysia, First edition, 2015,
  available at
  \url{https://www.cybersecurity.my/data/content_files/56/2073.pdf}. Accessed
  in May 2023.

\bibitem{sodiya2006towards}
A.~S. Sodiya, S.~A. Onashoga, and O.~Ajay{\~\i}, ``Towards building secure
  software systems.'' \emph{Issues in Informing Science \& Information
  Technology}, vol.~3, 2006.

\bibitem{daud2010secure}
M.~I. Daud, ``Secure software development model: A guide for secure software
  life cycle,'' in \emph{Proceedings of the International MultiConference of
  Engineers and Computer Scientists}, vol.~1, 2010, pp. 17--19.

\bibitem{farhan2018methodology}
A.~S. Farhan and G.~M. Mostafa, ``A methodology for enhancing software security
  during development processes,'' in \emph{Proceedings of the 21st Saudi
  Computer Society National Computer Conference}.\hskip 1em plus 0.5em minus
  0.4em\relax IEEE, 2018, pp. 1--6.

\bibitem{owaspsamm}
{OWASP}, ``{OWASP SAMM}, version 2,'' 2020, available at
  \url{https://owaspsamm.org/model/}. Accessed in May 2023.

\bibitem{bsimm}
``{BSIMM} trends \& insights,'' 2022, available at
  \url{https://www.synopsys.com/software-integrity/resources/analyst-reports/bsimm.html}.
  Accessed in May 2023.

\bibitem{weider2012towards}
D.~Y. Weider and K.~Le, ``Towards a secure software development lifecycle with
  {SQUARE+R},'' in \emph{Proceedings of the {IEEE} 36th Annual Computer
  Software and Applications Conference Workshops}, 2012, pp. 565--570.

\bibitem{van2017design}
A.~Van Den~Berghe, R.~Scandariato, K.~Yskout, and W.~Joosen, ``Design notations
  for secure software: a systematic literature review,'' \emph{Software \&
  Systems Modeling}, vol.~16, no.~3, pp. 809--831, 2017.

\bibitem{meland2008secure}
P.~H. Meland and J.~Jensen, ``Secure software design in practice,'' in
  \emph{Proceedings of the Third International Conference on Availability,
  Reliability and Security}.\hskip 1em plus 0.5em minus 0.4em\relax IEEE, 2008,
  pp. 1164--1171.

\bibitem{4359475}
C.~Haley, R.~Laney, J.~Moffett, and B.~Nuseibeh, ``Security requirements
  engineering: A framework for representation and analysis,'' \emph{IEEE
  Transactions on Software Engineering}, vol.~34, no.~1, pp. 133--153, 2008.

\bibitem{macdonald2016devsecops}
N.~MacDonald and I.~Head, ``{DevSecOps: How} to seamlessly integrate security
  into {DevOps},'' \emph{Gartner, Tech. Rep.}, 2016.

\bibitem{rajapakse2022challenges}
R.~N. Rajapakse, M.~Zahedi, M.~A. Babar, and H.~Shen, ``Challenges and
  solutions when adopting {DevSecOps: A} systematic review,'' \emph{Information
  and Software Technology}, vol. 141, p. 106700, 2022.

\bibitem{mohan2016secdevops}
V.~Mohan and L.~B. Othmane, ``Sec{D}ev{O}ps: Is it a marketing buzzword?
  {M}apping research on security in {D}ev{O}ps,'' in \emph{Proceedings of the
  11th international conference on availability, reliability and
  security}.\hskip 1em plus 0.5em minus 0.4em\relax IEEE, 2016, pp. 542--547.

\bibitem{farroha2014framework}
B.~S. Farroha and D.~L. Farroha, ``A framework for managing mission needs,
  compliance, and trust in the {DevOps} environment,'' in \emph{Proceedings of
  the {IEEE} Military Communications Conference}, 2014, pp. 288--293.

\bibitem{schneider2015security}
C.~Schneider, ``Security {DevOps} -- staying secure in agile projects,''
  \emph{OWASP AppSec Europe}, 2015.

\bibitem{rahman2016software}
A.~Rahman and L.~Williams, ``Software security in {DevOps: Synthesizing}
  practitioners’ perceptions and practices,'' in \emph{Proceedings of the
  {IEEE/ACM} International Workshop on Continuous Software Evolution and
  Delivery}, 2016, pp. 70--76.

\bibitem{cash2016managed}
S.~Cash, V.~Jain, L.~Jiang, A.~Karve, J.~Kidambi, M.~Lyons, T.~Mathews,
  S.~Mullen, M.~Mulsow, and N.~Patel, ``Managed infrastructure with {IBM} cloud
  {OpenStack} services,'' \emph{IBM Journal of Research and Development},
  vol.~60, no. 2-3, pp. 6--1, 2016.

\bibitem{de2014continuous}
S.~de~Vries, ``Continuous security testing in a {DevOps} world,'' \emph{OWASP
  AppSec Europe}, 2014.

\bibitem{myrbakken2017devsecops}
H.~Myrbakken and R.~Colomo-Palacios, ``Dev{S}ec{O}ps: a multivocal literature
  review,'' in \emph{Proceedings of the International Conference on Software
  Process Improvement and Capability Determination}.\hskip 1em plus 0.5em minus
  0.4em\relax Springer, 2017, pp. 17--29.

\bibitem{ahmed2019integrating}
Z.~Ahmed and S.~C. Francis, ``Integrating security with {DevSecOps: Techniques}
  and challenges,'' in \emph{Proceedings of the International Conference on
  Digitization}.\hskip 1em plus 0.5em minus 0.4em\relax IEEE, 2019, pp.
  178--182.

\bibitem{synopsis2020}
Synopsis, ``Dev{S}ec{O}ps practices and open source management in 2020,'' 2020,
  available at
  \url{https://www.synopsys.com/content/dam/synopsys/sig-assets/reports/rep-opensource-devsecops-survey-2020.pdf}.
  Accessed in May 2023.

\bibitem{AWSdevsecops}
S.~Manepalli, ``{AWS DevOps Blog.} {Building} end-to-end {AWS DevSecOps CI/CD}
  pipeline with open source {SCA, SAST and DAST} tools,'' Blog post, 2021,
  available at
  \url{https://aws.amazon.com/blogs/devops/building-end-to-end-aws-devsecops-ci-cd-pipeline\\-with-open-source-sca-sast-and-dast-tools/}.
  Accessed in May 2023.

\bibitem{mcgraw2006software}
G.~Mcgraw, ``Software security: Building security in,'' 2006.

\bibitem{payne2010}
J.~Payne, ``Integrating application security into software development,''
  \emph{IT Professional}, vol.~12, no.~2, pp. 6--9, 2010.

\bibitem{chakraborty2016}
M.~Chakraborty, ``Application security vs. software security: {What’s} the
  difference?'' Blog post, 2016, available at
  \url{https://www.synopsys.com/blogs/software-security/application-security-vs-software-security/}.
  Accessed in May 2023.

\bibitem{ISO27034}
ISO/IEC/IEEE, ``{ISO/IEC/IEEE} information technology — {S}ecurity techniques
  — {A}pplication security,'' \emph{ISO/IEC/IEEE 27034 First edition
  2011-11}, pp. 1--167, 2011, available at
  \url{https://www.iso.org/standard/44378.html}. Accessed in May 2023.

\bibitem{rebit2020}
M.~Nambiar, ``Re{BIT} application security framework,'' Tech. Rep., 2020,
  available at
  \url{https://pub.rebit.org.in/inline-files/ReBIT_Application_Security_Framework_2020.pdf}.
  Accessed in May 2023.

\bibitem{owaspappsec}
M.~Morana, T.~Gondrom, E.~Keary, A.~Lewis, S.~Tan, and C.~Watson, ``{OWASP}
  {Application} security guide for {CISOs},'' 2013, available at
  \url{https://owasp.org/www-pdf-archive/Owasp-ciso-guide.pdf}. Accessed in May
  2023.

\bibitem{grey}
J.~Tyndall, ``The {AACODS} checklist,'' 2010, available at
  \url{https://dspace.flinders.edu.au/xmlui/bitstream/handle/2328/3326/AACODS_Checklist.pdf?sequence=4}.
  Accessed in May 2023.

\bibitem{wohlin2014guidelines}
C.~Wohlin, ``Guidelines for snowballing in systematic literature studies and a
  replication in software engineering,'' in \emph{Proceedings of the 18th
  international conference on evaluation and assessment in software
  engineering}, 2014, pp. 1--10.

\bibitem{enisa2011}
ENISA, ``Secure software engineering initiatives,'' 2011, available at
  \url{https://www.enisa.europa.eu/publications/secure-software-engineering-initiatives}.
  Accessed in May 2023.

\bibitem{verdon2004risk}
D.~Verdon and G.~McGraw, ``Risk analysis in software design,'' \emph{IEEE
  Security \& Privacy}, vol.~2, no.~4, pp. 79--84, 2004.

\bibitem{web:lang:stats}
OWASP, ``Comprehensive, lightweight application security process,'' 2006,
  available at
  \url{https://owasp.org/www-pdf-archive/Us_owasp-clasp-v12-for-print-lulu.pdf}.
  Accessed in June 2023.

\bibitem{whitepapermicrosoft}
Microsoft, ``Security development lifecycle {SDL} process guidance,'' 2012,
  available at
  \url{https://www.microsoft.com/en-us/download/details.aspx?id=29884}.
  Accessed in June 2023.

\bibitem{safecode2018}
SAFECode, ``Fundamental practices for secure software development,'' 2018,
  available at
  \url{https://safecode.org/wp-content/uploads/2018/03/SAFECode_Fundamental_Practices_for_Secure_Software_Development_March_2018.pdf}.
  Accessed in June 2023.

\bibitem{adkins2020building}
H.~Adkins, B.~Beyer, P.~Blankinship, P.~Lewandowski, A.~Oprea, and
  A.~Stubblefield, \emph{Building Secure and Reliable Systems: Best Practices
  for Designing, Implementing, and Maintaining Systems}.\hskip 1em plus 0.5em
  minus 0.4em\relax O'Reilly Media, 2020.

\bibitem{bsimm12}
E.~Erlikhman, J.~Ewers, S.~Migues, and K.~Nassery, ``{BSIMM12},'' available at
  \url{https://www.bsimm.com/framework.html}. Accessed in June 2023.

\bibitem{NIST2016v1r1}
R.~Ross, M.~Winstead, and M.~McEvilley, ``Engineering trustworthy secure
  systems,'' National Institute of Standards and Technology, Tech. Rep., 2022,
  available at
  \url{https://nvlpubs.nist.gov/nistpubs/SpecialPublications/NIST.SP.800-160v1.pdf}.
  Accessed in June 2023.

\bibitem{jones2004secure}
R.~L. Jones and A.~Rastogi, ``Secure coding: building security into the
  software development life cycle,'' \emph{Inf. Secur. J. A Glob. Perspect.},
  vol.~13, no.~5, pp. 29--39, 2004.

\bibitem{sommerville2006}
I.~Sommerville, \emph{Software Engineering: (Update) (8th Edition)
  (International Computer Science)}.\hskip 1em plus 0.5em minus 0.4em\relax
  USA: Addison-Wesley Longman Publishing Co., 2006.

\bibitem{hoglund2004exploiting}
G.~Hoglund and G.~McGraw, \emph{Exploiting software: How to break code}.\hskip
  1em plus 0.5em minus 0.4em\relax Pearson Education India, 2004.

\bibitem{graham2006introduction}
D.~Graham, ``Introduction to the {CLASP} process,'' \emph{Build Security In},
  2006.

\bibitem{saltzer1975protection}
J.~H. Saltzer and M.~D. Schroeder, ``The protection of information in computer
  systems,'' \emph{Proceedings of the IEEE}, vol.~63, no.~9, pp. 1278--1308,
  1975.

\bibitem{safecode2017}
SAFECode, ``Tactical threat modeling,'' Tech. Rep., 2017, available at
  \url{https://safecode.org/wp-content/uploads/2017/05/SAFECode_TM_Whitepaper.pdf}.
  Accessed in June 2023.

\bibitem{thirdparty2017}
------, ``Managing security risks inherent in the use of third-party
  components,'' Tech. Rep., 2017, available at
  \url{https://safecode.org/wp-content/uploads/2017/05/SAFECode_TPC_Whitepaper.pdf}.
  Accessed in June 2023.

\bibitem{iso15288}
ISO/IEC/IEEE, ``{ISO/IEC/IEEE} international standard - systems and software
  engineering -- system life cycle processes,'' \emph{ISO/IEC/IEEE 15288 First
  edition 2015-05-15}, pp. 1--118, 2015.

\bibitem{NIST2021volume2}
R.~Ross, V.~Pillitteri, R.~Graubart, D.~Bodeau, and R.~McQuaid, ``Developing
  cyber resilient systems: A systems security engineering approach,'' National
  Institute of Standards and Technology, Tech. Rep., 2021, available at
  \url{https://nvlpubs.nist.gov/nistpubs/SpecialPublications/NIST.SP.800-160v2r1.pdf}.
  Accessed in June 2023.

\bibitem{arizon2021importance}
R.~Arizon-Peretz, I.~Hadar, and G.~Luria, ``The importance of security is in
  the eye of the beholder: Cultural, organizational, and personal factors
  affecting the implementation of security by design,'' \emph{IEEE Transactions
  on Software Engineering}, 2021.

\bibitem{dbir}
Verizon, ``Data breach investigations report {(DBIR)},'' 2022, available at
  \url{https://www.verizon.com/business/resources/reports/2022/dbir/2022-data-breach-investigations-report-dbir.pdf}.
  Accesses in June 2023.

\bibitem{spiekermann2018inside}
S.~Spiekermann, J.~Korunovska, and M.~Langheinrich, ``Inside the organization:
  Why privacy and security engineering is a challenge for engineers,''
  \emph{Proceedings of the IEEE}, vol. 107, no.~3, pp. 600--615, 2018.

\bibitem{alavi2014conceptual}
R.~Alavi, S.~Islam, and H.~Mouratidis, ``A conceptual framework to analyze
  human factors of information security management system {(ISMS)} in
  organizations,'' in \emph{Proceesings of the International Conference on
  Human Aspects of Information Security, Privacy, and Trust}.\hskip 1em plus
  0.5em minus 0.4em\relax Springer, 2014, pp. 297--305.

\bibitem{nist80053}
NIST, ``800-53 rev. 5: Security and privacy controls for information systems
  and organizations,'' Tech. Rep., 2020, available at
  \url{https://csrc.nist.gov/publications/detail/sp/800-53/rev-5/final}.
  Accessed in June 2023.

\bibitem{391832}
B.~Kitchenham, L.~Pickard, and S.~Pfleeger, ``Case studies for method and tool
  evaluation,'' \emph{IEEE Software}, vol.~12, no.~4, pp. 52--62, 1995.

\bibitem{georg2002using}
G.~Georg, I.~Ray, and R.~France, ``Using aspects to design a secure system,''
  in \emph{Proceedings of the {IEEE} International Conference on Engineering of
  Complex Computer Systems}, 2002, pp. 117--126.

\bibitem{saptrust}
J.~Schneider, ``New nist white paper on secure software development,'' 2020,
  available at
  \url{https://blogs.sap.com/2019/09/11/new-nist-white-paper-on-secure-software-development/}.
  Accessed in May 2023.

\bibitem{lipner2023inside}
S.~Lipner and M.~Howard, ``Inside the {Windows} security push: {A} twenty-year
  retrospective,'' \emph{IEEE Security \& Privacy}, 2023.

\bibitem{howard2003inside}
M.~Howard and S.~Lipner, ``Inside the {Windows} security push,'' \emph{IEEE
  Security \& Privacy}, vol.~1, no.~1, pp. 57--61, 2003.

\bibitem{busch2014evaluation}
M.~Busch, N.~Koch, and M.~Wirsing, ``Evaluation of engineering approaches in
  the secure software development life cycle,'' in \emph{Engineering Secure
  Future Internet Services and Systems}.\hskip 1em plus 0.5em minus 0.4em\relax
  Springer, 2014, pp. 234--265.

\bibitem{shostack2019empirical}
A.~Shostack, M.~Smith, S.~Weber, and M.~E. Zurko, ``Empirical evaluation of
  secure development processes,'' in \emph{Dagstuhl Reports}, vol.~9,
  no.~6.\hskip 1em plus 0.5em minus 0.4em\relax Schloss
  Dagstuhl-Leibniz-Zentrum fuer Informatik, 2019.

\bibitem{al2020maturity}
H.~Al-Matouq, S.~Mahmood, M.~Alshayeb, and M.~Niazi, ``A maturity model for
  secure software design: A multivocal study,'' \emph{IEEE Access}, vol.~8, pp.
  215\,758--215\,776, 2020.

\bibitem{khan2021systematic}
R.~A. Khan, S.~U. Khan, H.~U. Khan, and M.~Ilyas, ``Systematic mapping study on
  security approaches in secure software engineering,'' \emph{IEEE Access},
  vol.~9, pp. 19\,139--19\,160, 2021.

\bibitem{khan2022systematic}
------, ``Systematic literature review on security risks and its practices in
  secure software development,'' \emph{IEEE Access}, 2022.

\bibitem{acar2017developers}
Y.~Acar, C.~Stransky, D.~Wermke, C.~Weir, M.~L. Mazurek, and S.~Fahl,
  ``Developers need support, too: A survey of security advice for software
  developers,'' in \emph{Proceedings of the {IEEE} Cybersecurity Development
  (SecDev)}, 2017, pp. 22--26.

\bibitem{assal2019think}
H.~Assal and S.~Chiasson, ``'{T}hink secure from the beginning' a survey with
  software developers,'' in \emph{Proceedings of the 2019 {CHI} Conference on
  Human Factors in Computing Systems}, 2019, pp. 1--13.

\bibitem{sanchez2020security}
M.~S{\'a}nchez-Gord{\'o}n and R.~Colomo-Palacios, ``Security as culture: a
  systematic literature review of devsecops,'' in \emph{Proceedings of the
  {IEEE/ACM} 42nd International Conference on Software Engineering Workshops},
  2020, pp. 266--269.

\bibitem{stevanovic2011maturity}
B.~Stevanovi{\'c}, ``Maturity models in information security,''
  \emph{International Journal of Information and Communication Technology
  Research}, vol.~1, no.~2, 2011.

\bibitem{rabii2020information}
A.~Rabii, S.~Assoul, K.~O. Touhami, and O.~Roudies, ``Information and cyber
  security maturity models: a systematic literature review,'' \emph{Information
  \& Computer Security}, 2020.

\bibitem{le2016can}
N.~T. Le and D.~B. Hoang, ``Can maturity models support cyber security?'' in
  \emph{Proceedings of the {IEEE} International Performance Computing and
  Communications Conference}, 2016, pp. 1--7.

\bibitem{rea2017maturity}
A.~M. Rea-Guam{\'a}n, I.~S{\'a}nchez-Garc{\'\i}a, T.~San~Feliu, and
  J.~Calvo-Manzano, ``Maturity models in cybersecurity: A systematic review,''
  in \emph{Proceedings of the {IEEE} Iberian conference on information systems
  and technologies (CISTI)}, 2017, pp. 1--6.

\bibitem{uzunov2012engineering}
A.~Uzunov, E.~Fernandez, and K.~Falkner, ``Engineering security into
  distributed systems: A survey of methodologies,'' 2012.

\bibitem{suganya2018survey}
R.~Suganya, R.~A. Jothi, and V.~Palanisamy, ``A survey on security
  methodologies in e-voting system,'' \emph{International Journal of Pure and
  Applied Mathematics}, vol. 118, no.~8, pp. 511--515, 2018.

\bibitem{malik2012security}
A.~Malik and M.~M. Nazir, ``Security framework for cloud computing environment:
  A review,'' \emph{Journal of Emerging Trends in computing and information
  Sciences}, vol.~3, no.~3, pp. 390--394, 2012.

\bibitem{babar2011proposed}
S.~Babar, A.~Stango, N.~Prasad, J.~Sen, and R.~Prasad, ``Proposed embedded
  security framework for internet of things ({I}o{T}),'' in \emph{Proceedings
  of the {IEEE} International Conference on Wireless Communication, Vehicular
  Technology, Information Theory and Aerospace \& Electronic Systems
  Technology}, 2011, pp. 1--5.

\bibitem{pacheco2016iot}
J.~Pacheco and S.~Hariri, ``Io{T} security framework for smart cyber
  infrastructures,'' in \emph{Proceedings of the {IEEE} International workshops
  on Foundations and Applications of self* systems}, 2016, pp. 242--247.

\bibitem{enisa}
ENISA, ``Good practices for security of {I}o{T} - secure software development
  lifecycle,'' Tech. Rep., 2019, available at
  \url{https://www.enisa.europa.eu/publications/good-practices-for-security-of-iot-1}.
  Accessed in June 2023.

\bibitem{rindell2017busting}
K.~Rindell, S.~Hyrynsalmi, and V.~Lepp{\"a}nen, ``Busting a myth: Review of
  agile security engineering methods,'' in \emph{Proceedings of the 12th
  International Conference on Availability, Reliability and Security}, 2017,
  pp. 1--10.

\bibitem{kang2022cia}
S.~Kang and S.~Kim, ``{CIA}-level driven secure {SDLC} framework for
  integrating security into {SDLC} process,'' \emph{Journal of Ambient
  Intelligence and Humanized Computing}, vol.~13, no.~10, pp. 4601--4624, 2022.

\bibitem{ardo2022towards}
A.~A. Ardo, J.~M. Bass, and T.~Gaber, ``Towards secure agile software
  development process: A practice-based model,'' in \emph{Proceedings of the
  Euromicro Conference on Software Engineering and Advanced
  Applications}.\hskip 1em plus 0.5em minus 0.4em\relax IEEE, 2022, pp.
  149--156.

\bibitem{maher2020challenges}
Z.~Maher, A.~Shah, S.~Chan-dio, H.~Mohadis, and N.~Rahim, ``Challenges and
  limitations in secure software development adoption-a qualitative analysis in
  malaysian software industry prospect,'' \emph{Indian Journal of Science and
  Technology}, vol.~13, no.~26, pp. 2601--2608, 2020.

\bibitem{gasiba2020awareness}
T.~E. Gasiba, U.~Lechner, M.~Pinto-Albuquerque, and D.~M. Fernandez,
  ``Awareness of secure coding guidelines in the industry-a first data
  analysis,'' in \emph{Proceedings of the {IEEE} 19th International Conference
  on Trust, Security and Privacy in Computing and Communications}, 2020, pp.
  345--352.

\bibitem{kirlappos2013comply}
I.~Kirlappos, A.~Beautement, and M.~A. Sasse, ``“comply or die” is dead:
  Long live security-aware principal agents,'' in \emph{In proceedings of the
  International conference on financial cryptography and data security}.\hskip
  1em plus 0.5em minus 0.4em\relax Springer, 2013, pp. 70--82.

\bibitem{geer2010companies}
D.~Geer, ``Are companies actually using secure development life cycles?''
  \emph{Computer}, vol.~43, no.~6, pp. 12--16, 2010.

\end{thebibliography}

\end{document}